\documentclass[10pt, journal, letterpaper, table]{IEEEtran}
  
\usepackage{cite}
\usepackage{amsmath,amssymb,amsfonts}
\usepackage{algorithmic}
\usepackage{float}
\usepackage{graphicx}
\usepackage{mdframed}
\usepackage{textcomp}
\usepackage{tcolorbox}
\usepackage{cleveref} 

\usepackage{multirow}
\usepackage{booktabs}      
\usepackage{array}        
\usepackage{url}
\usepackage{ragged2e}      
\usepackage{listings}
\usepackage{xcolor}
\usepackage{threeparttable} 
\usepackage{svg}
\usepackage{pgfplots}
\usepackage{subcaption}

\pgfplotsset{width=10cm,compat=1.9}
\usepackage[nolist,nohyperlinks]{acronym}
\usepackage[normalem]{ulem}
\def\BibTeX{{\rm B\kern-.05em{\sc i\kern-.025em b}\kern-.08em
    T\kern-.1667em\lower.7ex\hbox{E}\kern-.125emX}}

\usepackage{boxedminipage}

\usepackage{soul}

\usepackage[export]{adjustbox}
\definecolor{nicolasred}{rgb}{0.7804, 0, 0.0431}
\definecolor{nicolascyan}{rgb}{0.1882, 0.7137, 0.7725}

\definecolor{ali}{rgb}{0.9, 0.1, 0.9}
\definecolor{done}{rgb}{0.298, 0.733, 0.090}

\tcbuselibrary{listings, skins} 
\definecolor{promptbackground}{rgb}{0.95, 0.95, 0.95}

\begin{document}

\title{Telco-oRAG: Optimizing Retrieval-augmented Generation for Telecom Queries via Hybrid Retrieval and Neural Routing}

\author{
 Andrei-Laurentiu Bornea$^*$,
 Fadhel Ayed$^*$,
 Antonio De Domenico$^*$,
 Nicola Piovesan$^*$,  Tareq Si Salem$^*$, 
 Ali Maatouk$^+$ \\
$^*$Paris Research Center, Huawei Technologies, Boulogne-Billancourt, France\\
$^+$Yale University, New Haven, Connecticut, USA 
\vspace{-1em}}

\maketitle

\begin{abstract}
Artificial intelligence will be one of the key pillars of the next generation of mobile networks (6G), as it is expected to provide novel added-value services and improve network performance. In this context, large language models have the potential to revolutionize the telecom landscape through intent comprehension, intelligent knowledge retrieval, coding proficiency, and cross-domain orchestration capabilities.
This paper presents Telco-oRAG, an open-source Retrieval-Augmented Generation (RAG) framework optimized for answering technical questions in the telecommunications domain, with a particular focus on 3GPP standards. Telco-oRAG introduces a hybrid retrieval strategy that combines 3GPP domain-specific retrieval with web search, supported by glossary-enhanced query refinement and a neural router for memory-efficient retrieval. Our results show that Telco-oRAG improves the accuracy in answering 3GPP-related questions by up to 17.6\% and achieves a 10.6\% improvement in lexicon queries compared to baselines. Furthermore, Telco-oRAG reduces memory usage by 45\% through targeted retrieval of relevant 3GPP series compared to baseline RAG, and enables open-source LLMs to reach GPT-4-level accuracy on telecom benchmarks.
\end{abstract}



\section{Introduction} 

Modern mobile networks are among the most complex engineered systems in operation today. They must support a wide range of services, spanning enhanced mobile broadband, ultra-reliable low-latency communication, and massive machine-type communication, while maintaining compliance with intricate and evolving technical standards such as those defined by the \ac{3GPP}. The ongoing transition toward 5G-Advanced and early visions of 6G further compound this complexity, introducing increasingly dense deployments, heterogeneous network components, and stringent performance requirements.

To manage this complexity at scale, future networks are expected to evolve into AI-native infrastructures, where \ac{AI} systems assist or automate critical tasks across network planning, configuration, optimization, and fault resolution. Within this broader vision, \acp{LLM} are gaining traction as enablers of intent-based interfaces, explainable automation, and intelligent support systems.

However, vanilla \acp{LLM} rely solely on their internal representations and learned parameters to generate text, and they struggle in specialized domains such as telecommunications, where queries often involve domain-specific terminology, implicit standards knowledge, and subtle contextual dependencies. Even advanced models like GPT-4 exhibit limited reliability when tasked with answering questions related to 3GPP specifications~\cite{maatouk2023teleqna}. Addressing these limitations is essential to fully realize the promise of AI-native networks.

In this paper, we introduce Telco-oRAG, a framework specifically designed to enhance the \ac{LLM} understanding and knowledge of the telecommunication domain. Such framework is not limited to chatbot applications, but is broadly applicable to any telecom task requiring accurate interpretation of telecom standards, representing a step forward toward practical AI-native network operations.

\subsection{Related Works}\label{s:related_work}

\ac{LLM} capabilities have attracted notable attention in diverse industrial domains, including highly specialized fields such as telecommunications. In general, two complementary strategies have emerged to adapt LLMs to domain-specific tasks: (1) enhancing model knowledge through domain-specific fine-tuning, and (2) incorporating external knowledge at inference time via \ac{RAG}.


Fine-tuning offers an effective means to inject domain expertise into \acp{LLM} \cite{gururangan2020don}. However, it requires substantial computational resources and is prone to overfitting when trained on limited data \cite{thompson2020computational}. Moreover, fine-tuning exhibits reduced flexibility in rapidly evolving domains, which leads \ac{LLM} vendors to periodically retrain their models to keep up with updates in specialized domains \cite{IBM}. This makes any LLM obsolete a few months after it is released and results in unstable performance from one release to another \cite{chen2023chatgptsbehaviorchangingtime}, making \ac{LLM} benchmarking very challenging.



Recent research has explored parameter-efficient strategies to mitigate the high computational costs of fine-tuning \acp{LLM} for domain-specific tasks \cite{zhang2024instructiontuninglargelanguage}. Conventional fine-tuning requires storing large gradient states and domain-specific representations, leading to significant memory requirements. Techniques such as adapter layers~\cite{Mahabadi2021}, prefix-tuning \cite{li-liang-2021-prefix}, low-rank adaptation (LoRA)~\cite{hu2021lora}, or combining low-rank updates with quantization of the frozen base model (QLoRA)~\cite{Dettmers2023}, have been proposed to minimize the number of trainable parameters while preserving model performance. Importantly, a late work has shown that LoRA does not inject sufficient additional knowledge to adapt the \acp{LLM} to the telecommunications field~\cite{maatouk2024telellmsseriesspecializedlarge}.

Recent efforts in telecom-specific fine-tuning include TelecomGPT~\cite{zou2024telecomgpt}, which leverages continual pre-training, instruct tuning with \ac{SFT}, and alignment tuning with \ac{DPO} \cite{rafailov2024directpreferenceoptimizationlanguage}. Each of these phases uses a specifically designed telecom dataset constructed using on public documents.
Moreover, Tele-LLMs is a series of \acp{LLM} \cite{maatouk2024telellmsseriesspecializedlarge}, ranging from 1B to 8B parameters developed using continual pre-training on a dedicated telecom dataset (Tele-Data). 

Although these work have shown notable results, they inherits the computational burdens of fine-tuning: First, acquiring high-quality, annotated telecom-specific datasets suitable for fine-tuning is costly and time-consuming. Second, the fast-paced evolution of standards (e.g., 3GPP Releases) necessitates frequent retraining to ensure model relevance. Finally, fine-tuning methods often entangle learned knowledge within model weights, making it difficult to update or remove outdated information, a limitation particularly critical for regulatory and compliance driven sectors like telecom. 

\ac{RAG} has emerged as an alternative pathway by decoupling the injection of knowledge from the training of model parameters. By retrieving the relevant context from external knowledge sources at inference time, \ac{RAG} reduces reliance on static model parameters and allows for more efficient adaptation to new information~\cite{Lewis2020NIPS}. This approach has proven particularly effective in knowledge-intensive tasks, improving factual accuracy while avoiding the computational costs of periodic fine-tuning~\cite{izacard2022few}.

In recent years, the research community has proposed several advancements to improve \ac{RAG}, which can be classified as follows \cite{zhao2024retrievalaugmentedgenerationaigeneratedcontent}: 1) input enhancement, including query transformation and data augmentation; 2) retriever enhancement, such as chunk optimization, retriever finetuning, hybrid retrieval, and re-ranking; 3) generator enhancement, which includes prompt engineering, decoding tuning, and generator finetuning; 4) result enhancement, which allows \ac{RAG} output rewrite; 5) RAG pipeline enhancement, including adaptive retrieval and iterative \ac{RAG}. 

The integration of \ac{RAG} into telecommunications systems has shown substantial value~\cite{piovesan2024telecom}. Indeed, telecommunications systems support tools such as chatbots that streamline the access of professionals to standards, accelerating research, development and improving regulatory compliance. Few \ac{RAG} frameworks have been developed to address the complexities of technical standards and their rapid evolution. For instance, {TelecomRAG}~\cite{Yilma2025} utilizes a knowledge base built from 3GPP Release 16 and Release 18 specification documents to provide responses to telecom standard query. Importantly, in {TelecomRAG}, the retriever transform each new query using the history of past queries and responses.
{Telco-RAG}~\cite{bornea2024telcorag} is an open-source framework designed to handle the specific needs of queries about telecommunications standards, by integrating domain-specific input, retrieval, and generator enhancements. Additionally, {Chat3GPP}~\cite{huang2025chat3gpp} offers another open-source \ac{RAG} tailored for 3GPP specifications, combining chunking strategies, 
and re-ranking. More recently, researchers have proposed CommGPT \cite{jiang2025}, a telecom specialized \ac{LLM} constructed using dedicated continuing pre-training dataset and instruction fine-tuning dataset. The fine-tuned model is then combined with \ac{KG} \cite{abusalih2021domainspecificknowledgegraphssurvey} and \ac{RAG} to assist CommGPT in generating more precise and comprehensive responses. Although these approaches introduce significant novelties, leading to notable results, they all lack a means to simultaneously provide \acp{LLM} 1) a macroscopical vision of the topics related to the user query, 2) up-to-date information, and 3) precise understanding of domain-specific technical terms and abbreviations.

\subsection{Main Contributions}\label{s:contributions}
\textit{What are the main research challenges to be addressed when designing an \ac{LLM}-based chatbot for telecom queries? }\textit{How to make and maintain the chatbot accurate and simultaneously resource efficient and inexpensive? }\textit{What are the hyperparameters that should be carefully tuned to optimize the chatbot performance? } \textit{How to use a transparent evaluation, enhance reproducibility, and support future research?} 

These are a few of the questions that we tackle with our contributions, which are as follows:

\begin{enumerate}
    \item We present Telco-oRAG, a novel \ac{RAG} pipeline designed to answer queries on telecommunication networks, and in particular to address challenging standard queries. We prove that the proposed pipeline is effective across \acp{LLM} of different sizes and that it helps mid-sized \acp{LLM} to perform closely to proprietary \acp{LLM} on domain-specific knowledge. Notably, we show that Telco-oRAG outperforms existing \ac{LLM} specialized on telecom domain.

    \item We study \ac{RAG} input enhancement, retriever enhancement, generator enhancement, and pipeline enhancement. We highlight the impact of key RAG parameters and show that the optimal hyperparameter setting yields a 17.6\% accuracy gain over vanilla \acp{LLM}.

    \item We develop a hybrid retriever that complements the context extracted from 3GPP documents with data selected from the web. Our results highlight that web search is key to provide accurate answers to standard overview queries. Moreover, we design a dual rounds retriever where the second round of retrieval leverages a query augmented by the output of the first round of retrieval. The dual rounds retriever increases the accuracy of the baseline retriever on standard query of about 2$\%$.

    \item We design an \ac{NN} router that selects the most appropriate sources of information for answering user queries. This approach makes Telco-oRAG scalable with respect to future knowledge bases describing new technologies and products. Our results show that the proposed \ac{NN} router is more accurate than off-the-shelf classifiers and that it reduces memory usage of baseline \ac{RAG} by 45\%. Moreover, we analyse the failure cases of the proposed NN router and their impact on the overall performance.

     \item We provide an end-to-end latency analysis of Telco-oRAG, breaking down per-stage costs and discussing latency-performance tradeoff of each component in Telco-oRAG.

     \item We extend our evaluation to open-ended QnA using the LLM-as-judge method. These results confirm the value of Telco-oRAG that leads up to 42.8 percentage-points gains with respect to baseline models.

    \item We have made Telco-oRAG available as an open-source chatbot\footnote{\url{https://github.com/netop-team/Telco-RAG}} together with the dataset used for its evaluation.\footnote{\url{https://github.com/netop-team/TeleQnA}} This effort is expected to significantly contribute to future research efforts on large \ac{AI} models for future wireless communication systems.
    
\end{enumerate}
 This journal significantly extends our earlier conference manuscript~\cite{bornea2024telcorag}:  we introduce (i) an improved query refinement, (ii) a web retrieval module, (iii) a methodological optimization of the \ac{RAG} hyperparameters, and (iv) an extensive set of numerical evaluations including latency analysis and performance evaluation on open-ended questions. 

 The remainder of this paper is structured as follows. We provide an introduction to \ac{RAG} in Section~\ref{sec:RAG}. The Telco-RAG pipeline is then presented in Section~\ref{sec:methodology}. Experimental evaluation of our proposed RAG is detailed in Section~\ref{section:experiments}. Finally, we conclude the paper and outline potential avenues for future research in Section~\ref{s:conclusion}.

\begin{figure*}[ht]
\centering
\includegraphics[width=1.\textwidth]{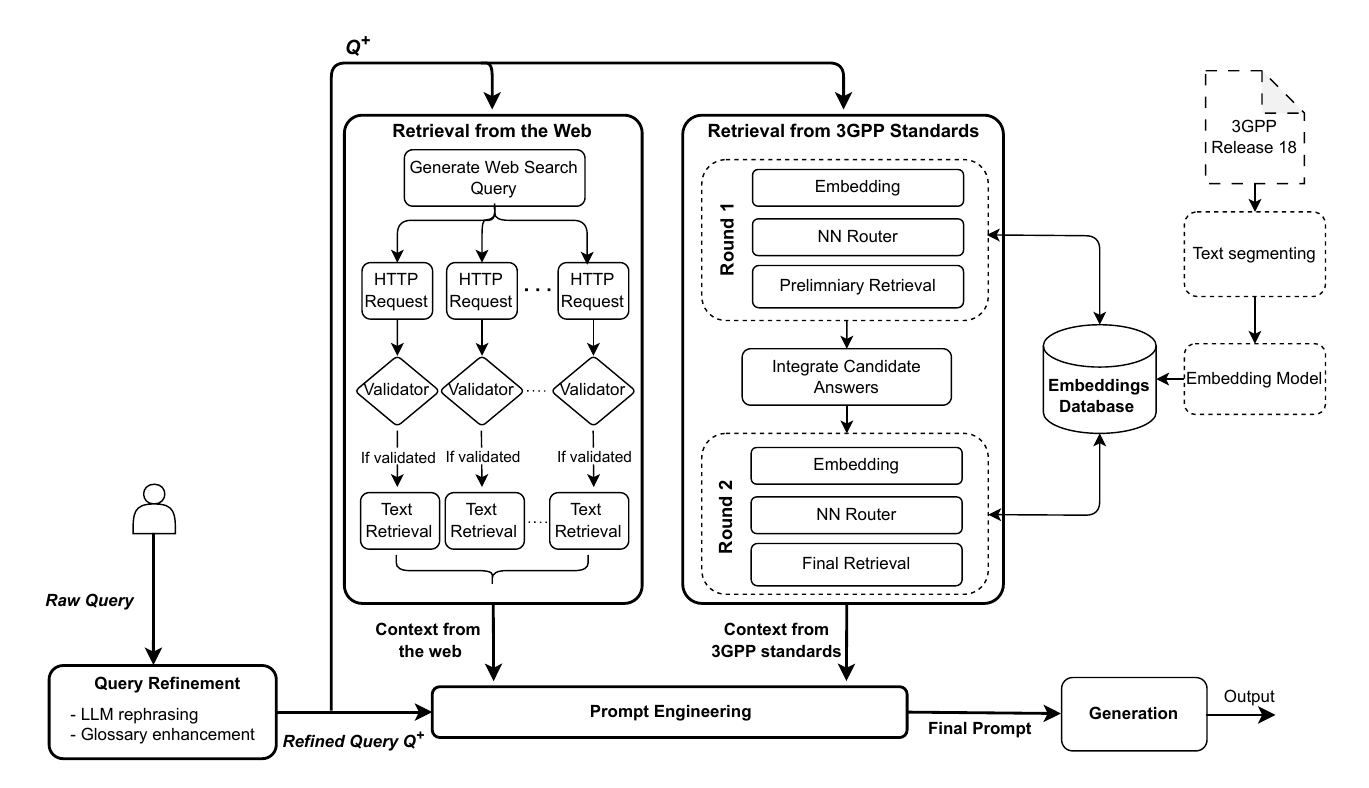}
\caption{The proposed Telco-oRAG pipeline. The left side shows the web search block; the right side shows the content retrieval from the 3GPP database.} 


\label{fig:pipeline}
\end{figure*}


\section{Preliminaries on RAG}
\label{sec:RAG}

\ac{RAG} is a framework that enhances the response quality of a \ac{LLM} by incorporating relevant external knowledge retrieved from a document corpus at inference time. This is especially useful for domain-specific applications such as telecommunications, where pre-trained models may lack access to evolving standards or specialized terminology.

Let $\mathcal{D} $ denote a large corpus of unstructured documents, and let $\mathcal{C} = \{ c_1, c_2, \dots, c_N \} $ be the set of document segments (or chunks) generated by partitioning $\mathcal{D} $ into fixed-length, non-overlapping  units. In \ac{RAG}, each chunk $c_i \in \mathcal{C} $ is transformed into a dense vector representation $\mathbf{v}_i \in \mathbb{R}^d $ using an embedding function:
\begin{align*}
f_{\text{embed}} : \mathcal{C} \rightarrow \mathbb{R}^d, \quad \mathbf{v}_i = f_{\text{embed}}(c_i).
\end{align*}

The embedding model---e.g., OpenAI’s \emph{text-embedding-3-large}~\cite{openai2023embeddings}---encodes the semantic content of each chunk into a continuous vector space, enabling efficient similarity-based retrieval.

At inference time, a user query $q \in \mathcal{Q} $ is embedded using the same function:
\begin{align*}
\mathbf{v}_q = f_{\text{embed}}(q).
\end{align*}
To retrieve relevant information, \ac{RAG} compares $\mathbf{v}_q $ against the chunk embeddings $\{ \mathbf{v}_i \} $ via cosine similarity:
\begin{align*}
\text{sim}(\mathbf{v}_q, \mathbf{v}_i) = \frac{\mathbf{v}_q \cdot \mathbf{v}_i}{\| \mathbf{v}_q \| \cdot \| \mathbf{v}_i \|}.
\end{align*}

Since all embeddings are normalized ($\| \mathbf{v}_q \| = \| \mathbf{v}_i \| = 1 $), this metric becomes equivalent to the inner product, thus maximizing cosine similarity is equivalent to minimizing the Euclidean distance.
Then, \ac{RAG} selects the top-$k $ most relevant chunks to form the retrieval set:
\begin{align*}
\mathcal{C}_{\text{top-}k}(q) = \underset{c_i \in \mathcal{C}}{\text{Top-}k} \ \text{sim}(\mathbf{v}_q, \mathbf{v}_i).
\end{align*}
Then, $\mathcal{C}_{\text{top-}k}(q) $ is concatenated with the query and passed as input to the language model to generate the final output.

\ac{RAG} decouples retrieval from generation; thus, new or updated knowledge can be incorporated by simply updating the document corpus and its corresponding embeddings, without requiring any retraining or fine-tuning of the LLM.

In the next section, we introduce the proposed Telco-oRAG, a \ac{RAG} framework tailored to the unique structure and language of telecom-standard documents.

The folowing discussion mainly focuses on answering to questions related to 3GPP
questions. This choice was done considering the importance of 3GPP in the telecommuni-
cation domain. However, in our results, we show that even a Telco-oRAG implementation
targeting 3GPP standards is able to answer to questions from other sources; moreover,
extending this pipeline on other standards or documents does not need any architecture
change and has as its main focus 3GPP Standards, both as questions and documents.

This was only a matter of choice, driven by the underperfomance of LLMs on this set
of questions. Future research can test the full Telco-oRAG pipeling on other standards
or documents without any arhitectural modifications. It only requires: (i) updating the
embeddings database, (ii) updating abbreviations and technical terms, and (iii) updating
the neural router training set (for instance considering the different ITU recommendations
series) and its model accordingly.

\section{Telco-oRAG}\label{sec:methodology}
Telco-oRAG is designed to address the unique challenges of telecom-related queries, which often involve ambiguous terminology, layered technical references, and evolving documentation. As illustrated in the architecture shown in Fig.~\ref{fig:pipeline}, Telco-oRAG produces accurate, context-aware answers by integrating information from two complementary sources: structured content from 3GPP documents and data from the web.

Before retrieving any content, the pipeline begins with a {query refinement} stage (presented in Section \ref{sec:Query Augmentation}). In this stage, the raw user query is first rephrased using a language model to clarify intent and improve readability. Next, telecom-specific glossaries are used to expand abbreviations and technical terms, producing an enriched query that captures domain-specific nuances.

With this enhanced query, Telco-oRAG launches a hybrid retrieval processes. The {web search} pipeline (see Section \ref{subsec:webscraper}) issues web queries, and uses an LLM to iteratively validate and filter returned content from the web.
In parallel, the {retrieval} pipeline (described in Section \ref{subsec:RAGRAM}) performs a dual rounds retrieval from 3GPP documents, guided by a neural network-based router and augmented by candidate answers generated by an intermediate LLM.

Finally, the {prompt engineering} block (see Section \ref{subsec:prompt}) assembles the retrieved web content, selected 3GPP passages, and refined query into a structured prompt, which is used to generate a grounded and precise LLM response.

In the following, we will detail each block of Telco-oRAG.

\subsection{User Query Refinement}
\label{sec:Query Augmentation}

User queries related to telecommunication standards often suffer from two primary challenges: (1) the high density of domain-specific technical terms and abbreviations, and (2) the difficulty for \ac{RAG} systems to accurately infer user intent. These factors frequently result in the retrieval of semantically related, yet contextually irrelevant, information.






For a given input sentence $s $, the corresponding embedding $\mathbf{v}_s = f_{\text{embed}}(s) $ is computed based on the distribution of its constituent tokens, which may include technical terms or abbreviations. However, when such terms appear without sufficient contextual information, the resulting embedding $\mathbf{v}_s $ may fail to accurately capture their intended semantics. For example:
\begin{align*}
s = \text{``What role does PCRF play in QoS control?"},
\end{align*}
in which the term ``PCRF'' lacks contextual grounding. As a result, the embedding may poorly align with relevant standard documents, reducing the retrieval effectiveness.

In the following we present the methodology designed to address this challenge:

\subsubsection{LLM-rephrasing}
\label{subsubsec:llm-rephrase}

First, the raw query is rephrased by a LLM to produce a clearer and grammatically correct query, denoted by $\mathcal{Q} $.
Although the semantic content of the original query is preserved, this rephrasing helps to eliminate ambiguities caused by informal phrasing, typographical errors, or incomplete sentences. In practice, this step improves the quality of the generated embedding, even before any domain-specific enhancement is applied.

\subsubsection{Glossary-enhancement}

To improve the semantic representation of telecom-related queries in the embedding space, 
domain-specific clarifications for abbreviations and technical terms are incorporated to the LLM-rephrased query, $\mathcal{Q} $. 

To achieve this goal, we construct two domain dictionaries based on the ``Vocabulary for 3GPP Specifications''~\cite{3GPPTR21.905}: 
\begin{itemize}
    \item $\mathcal{D}_\text{abbr}(\cdot) $ maps known abbreviations to their full expansions.
    \item $\mathcal{D}_\text{terms}(\cdot) $ maps technical terms to their standard definitions.
\end{itemize}

Let $\{A_i\}_{i=1}^m $ and $\{T_j\}_{j=1}^n $ represent the abbreviations and technical terms identified within the LLM-rephrased query $\mathcal{Q} $, respectively. The refined query $\mathcal{Q}^+ $ is defined as:
\begin{align*}
\textstyle\mathcal{Q}^+ = \mathcal{Q} \cup \big( \bigcup_{i=1}^{m} \mathcal{D}_\text{abbr}(A_i) \cup \bigcup_{j=1}^{n} \mathcal{D}_\text{terms}(T_j) \big).
\end{align*}
By appending abbreviations expansions and technical terms definitions to the query, the resulting embedding better captures the technical semantics, leading to improved alignment with relevant documents during retrieval.

An illustration of the query refinement process can be found in Box  1.

\begin{tcolorbox}[colframe=black!75!white, colback=gray!10!white, title=
Box 1. Illustration of Query Refinement
]
    \begin{footnotesize}
        \textbf{Raw Query:} ``Why is the association pattern period for PRACH introduced in NR, and why is it needed?" 

\textbf{LLM-rephrased Query} ``What is the purpose of introducing the association pattern period for PRACH in NR (New Radio) standards?" 

\textbf{Terms and Definitions:} 
\begin{itemize}
    \item \textbf{NR}: Fifth generation radio access technology.
    \item \textbf{PRACH}: Physical Random Access Channel.
    \item \textbf{Association Pattern Period}: Defines the interval in which a specific access pattern repeats.
\end{itemize}
    \end{footnotesize}
\label{box:examplequery}
\end{tcolorbox}

\subsection{Retrieval from the Web}\label{subsec:webscraper}

As shown in Fig.~\ref{fig:pipeline}, Telco-oRAG integrates a web information retrieval module designed to fetch relevant online data in real time. This module handles both I/O-bound and CPU-bound operations using asynchronous and parallel programming paradigms, respectively.

Web retrieval begins by submitting the refined query $\mathcal{Q}^+ $ to public search engines. The returned results contain ranked URLs along with short text fragments---or snippets---which highlight the relevance of each URL to the query. Telco-oRAG uses the snippets generated by the search engines as anchor points: for each, we extract a 250-token paragraph centered around the snippet location to capture sufficient information.

The I/O-bound portion of this workflow, which includes issuing HTTP requests and downloading documents of different types (e.g., HTML and PDF), is managed using asynchronous, non-blocking execution\footnote{A non-blocking request allows the system to initiate an I/O operation---such as an HTTP request---without halting the program execution while waiting for the response. In this way, the system can continue executing other tasks in parallel, improving overall responsiveness and resource utilization.}. Although Telco-oRAG handles one user query at a time, this mechanism enables to retrieve multiple documents in parallel, significantly reducing inference time.

Once the content is retrieved, Telco-oRAG performs CPU-bound processing steps such as document parsing, text cleaning, and semantic validation. These operations are parallelized across multiple CPU cores to maintain scalability. Of particular importance is the validation step, which determines whether a given paragraph is relevant to the query. 

\paragraph*{LLM-based Snippet Validation} 
We implement an LLM-based validator module that processes retrieved online paragraphs in batches to select the most relevant sources for the context. For each batch, the LLM classifies every paragraph either as relevant (``True'') or irrelevant (``False'') with respect to the user query. Paragraphs marked ``True'' are retained for inclusion in the final context prompt.

To avoid 
selecting more relevant paragraph than can fit within the context window of the LLM---and thereby wasting computing resources---we 
employ a sequential batch validation strategy with early stopping. Specifically, the system initially retrieves a large set of candidate paragraphs and evaluates them in batches. As soon as the number of relevant paragraphs reaches a predefined threshold (determined by the LLM context budget), the validation loop halts. 

\subsection{Retrieval from 3GPP standards}
\label{subsec:RAGRAM}
\subsubsection{Dual rounds retrieval}
\label{subsubsec:candidate-integration}
Telco-oRAG performs the offline retrieval from the 3GPP documents using a dual rounds approach. 
In the first round, the refined query $\mathcal{Q}^+ $ is used to retrieve an initial set of relevant passages from the 3GPP corpus. 
These passages are then provided to an \ac{LLM}, which generates a set of $k$ candidate answers, denoted by $\mathcal{A} = \{a_1, a_2, \dots, a_k\} $. 

The candidate answers serve two purposes: they help refine the interpretation of the query and offer preliminary hypotheses about potential answers, thereby guiding the second retrieval round toward more targeted and relevant content.

In the second round, an augmented query $\mathcal{Q}^{++} = \mathcal{Q}^{+} \cup \mathcal{A} $ is constructed by appending the candidate answers to the refined query. This further augmented query is then used to perform a more targeted retrieval from the 3GPP corpus.

This dual rounds design is especially valuable in highly technical domains such as telecommunication standards, where initial queries are often ambiguous, and a single retrieval may fail to surface the most relevant information. By combining an initial query refinement stage (see Section \ref{sec:Query Augmentation}) with augmentation through model-generated candidate answers, Telco-oRAG significantly improves alignment between the original query and the retrieved content.

\begin{figure}[t]
\centering
\includegraphics[width=\linewidth]{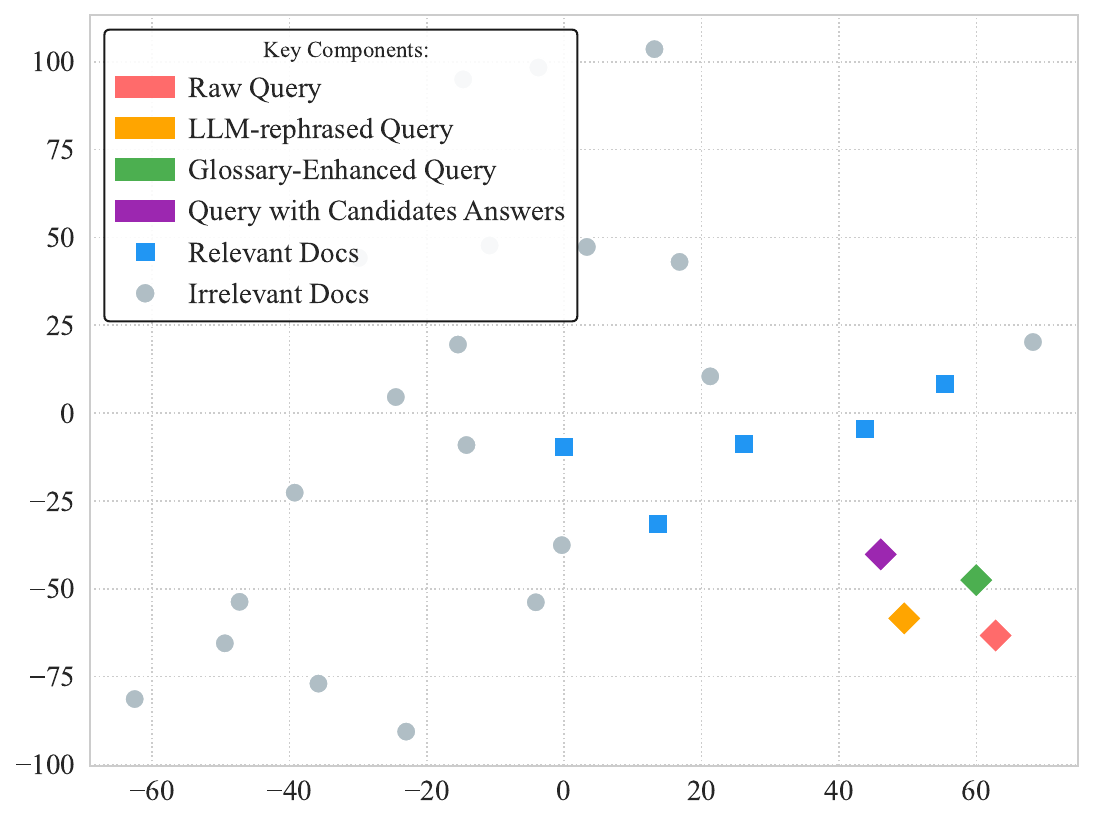}
\caption{t-SNE projection of the embeddings of 3GPP documents and user query after the proposed processing stages.
}
\label{fig:tsne-query-augmentation}
\end{figure}

The effectiveness of this approach is illustrated in Fig.~\ref{fig:tsne-query-augmentation}, where t-distributed stochastic neighbor embedding (t-SNE) projections show how each successive stages progressively improve the alignment between the query embedding and the embeddings of the relevant 3GPP documents.

\subsubsection{Reducing the RAM requirements of Telco-oRAG}
\label{sec: RAG's RAM usage}

Our experiments have shown that processing small chunks of text can improve \ac{LLM} performance;
however, reducing chunk size significantly increases memory requirements. Let $L $ denote the total number of tokens in the corpus $\mathcal{D} $, and $\ell $ the chunk size (in tokens). 
Each chunk $c_i \in \mathcal{C} $ is embedded into a fixed-dimensional space $\mathbb{R}^d $ via $f_{\text{embed}} : \mathcal{C} \to \mathbb{R}^d $, yielding total memory cost:
\begin{align*}
\text{Memory} \propto 
\left\lceil {L}/{\ell} \right\rceil \cdot d,
\end{align*}
where $d $ is the embedding dimension.
Since $d $ is constant (e.g., 1024 for \texttt{text-embedding-3-large}), smaller $\ell $ leads to a larger number of chunks in the embedding database and to a linear growth in RAM usage. For example, embedding all 3GPP Release~18 documents with $\ell = 125 $ requires approximately 11.5\,GB of RAM.
 
To address this issue, we recall that the \ac{3GPP} standards categorize specifications into 18 distinct series numbered from 21 to 38~\cite{specsbyseries}. Each series provides the technical details of a specific aspect of \ac{3GPP} standards (e.g., radio access, core network components, security). To filter out irrelevant information and reduce the \ac{RAG} requirements on \ac{RAM} resources, we have developed an \ac{NN} router tailored to infer the \ac{3GPP} series that contains required information for providing the correct answer to the user queries. 

 \begin{figure}[ht]
\centering
\includegraphics[width=1\linewidth]{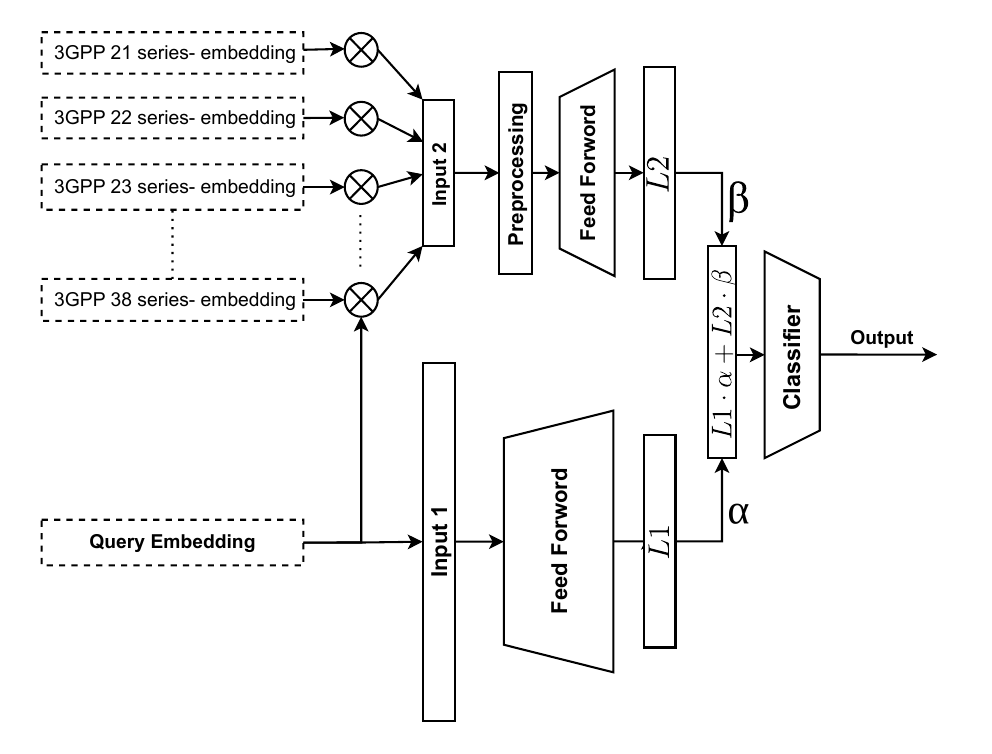}
\caption{The proposed NN router architecture.} 
\label{fig:tsneplot}
\end{figure}

The architecture of the proposed \ac{NN} router is illustrated in Fig.~\ref{fig:tsneplot}.
The process begins by constructing a high-level summary for each of the 18 \ac{3GPP} specification series~\cite{specsbyseries}, using a dedicated \ac{LLM}. These summaries are short textual descriptions that capture the thematic content of each series (e.g., radio protocols, access, core network). An example of one of the summaries, the one related to the Requirements series, is depicted in Box 2. Each summary is then embedded using the \texttt{text-embedding-3-large} model, resulting in a 1024-dimensional representation per series.

The NN router takes two inputs: 
\begin{itemize}
    \item \textbf{Input 1:} A 1024-dimensional embedding vector representing the refined query $Q^+ $, obtained via the same embedding model used for the summary.
    \item \textbf{Input 2:} An 18-dimensional vector where each entry is the inner product between the query embedding and the embedding of a corresponding series summary.
\end{itemize}

These two input channels provide complementary information: global semantic context from the query embedding (input 1), and relative alignment scores across all \ac{3GPP} series (input 2). The \ac{NN} router combines these signals to output a binary vector indicating which series are likely to contain relevant content for the query.

To process the query embedding, the network uses a few linear layers to reduce its size from 1024 to 256 dimensions. This path also includes dropout to prevent overfitting and batch normalization to keep training stable. In parallel, the second input---the 18-dimensional vector of alignment scores---is first passed through a softmax layer and then projected up to 256 dimensions so that it can be combined with the query branch.

The outputs from both branches are then combined using two trainable scalar weights, $\alpha $ and $\beta $, which modulate the contribution of each input stream to the final prediction. The resulting 256-dimensional joint representation is passed through a classification head to generate a probability vector, which represents the relevance of the 18 3GPP series with respect to the user query. Finally, the k-most relevant series are selected and the related content is loaded in memory.

To train the NN router, we constructed a dedicated dataset of 30,000 questions extracted from 500 documents in 3GPP Release 18, with each question labeled by its originating series. This approach avoids overfitting and ensures a robust evaluation of the Telco-oRAG pipeline~\cite{Gilardi2023}.
 \begin{tcolorbox}[colframe=black!75!white, colback=gray!10!white, title=Box 2. Series 21: Requirements]
    \textbf{Summary:} Requirements (\ac{3GPP} 21 series) focuses on the overarching requirements necessary for UMTS (Universal Mobile Telecommunications System) and subsequent cellular standards. This series addresses enhancements to GSM, establishes foundational security standards, and provides guidance on the general evolution of 3GPP systems, offering a cohesive set of requirements that shaped both UMTS and the continued development of mobile communications standards.
\end{tcolorbox}

\subsection{Prompt Engineering}
\label{subsec:prompt}

Prompt formulation plays a critical role in RAG, as it determines how effectively the language model incorporates retrieved context to generate a correct and concise answer~\cite{CHEN2025101260}. In this work, we design a structured prompt format optimized for clarity and alignment with the user query.

The final prompt in Telco-oRAG follows a dialogue-oriented structure, which has been shown to improve LLM performance in multi-hop and context-heavy reasoning tasks~\cite{Gao2019NeuralAT}. The prompt begins with the refined query $\mathcal{Q}^{+} $, followed by the validated context retrieved from both 3GPP specifications and web documents.

To ensure the model remains focused on the question, we repeat the query just before presenting the answer options.\footnote{In order to design a general \ac{RAG} framework that can be used both for open-ended a multiple-choice questions, the answer options are not used in the retrieval stage of Telco-oRAG and only included in the final prompt.} This repetition acts as a final anchor, reinforcing the task instruction and mitigating the risk of the LLM drifting off-topic during generation.

The full prompt structure is shown in Box 3.

 \begin{tcolorbox}[colframe=black!75!white, colback=gray!10!white, title= Box 3. Final Prompt]
                  
*Please provide the answer to the following question: \textless User Query\textgreater\\
        *Terms and Definitions: \textless Defined Terms\textgreater\\
        *Abbreviations: \textless Abbreviations\textgreater\\
        *Considering the following context:
        \textless Retrieved Context\textgreater\\
        *Please provide the answer to the following question: \textless User Query\textgreater
\end{tcolorbox}

\begin{figure*}[ht]
\centering
\includegraphics[width=1\linewidth, frame]{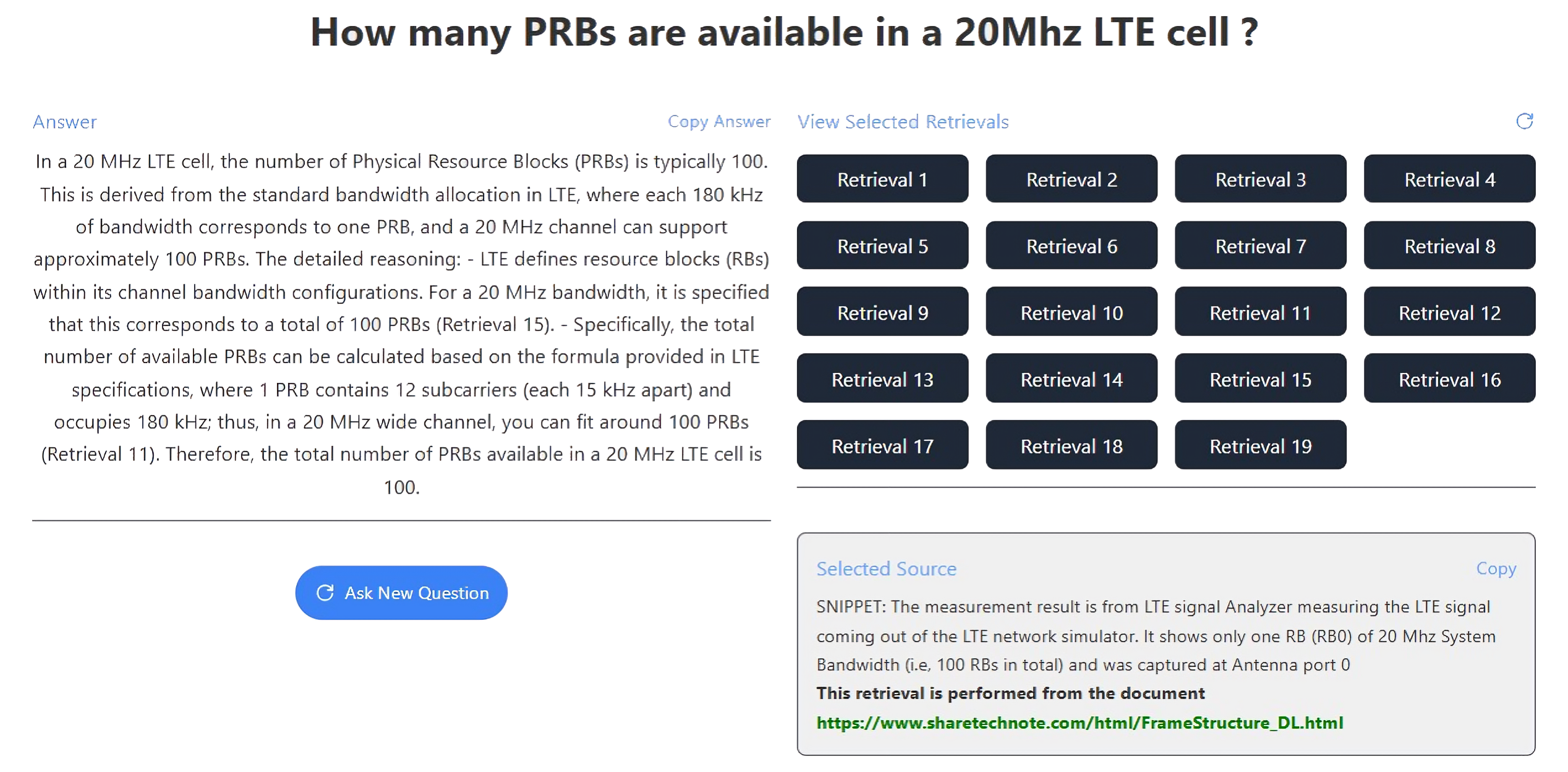}
\caption{Telco-oRAG frontend. Left side: the answer provided by Telco-oRAG to the user query.  Right side: the selected retrievals that support the generation of the answer.} 

\label{fig:frontend}
\end{figure*}

Before concluding this section, we show in Figure \ref{fig:frontend} the frontend of Telco-oRAG and the answer provided to an open question related to the 3GPP standard. On the right side, Telco-oRAG reports the selected retrievals, and in the bottom-right corner, we highlight one snippet retrieved from the web.

\section{Experimental Results}\label{section:experiments}

In this section, we evaluate the performance of Telco-oRAG in answering questions related to telecommunications standards. To assess the effectiveness of our approach, we compare Telco-oRAG against several benchmarks, including current state-of-the-art \acp{LLM} and Telco-RAG \cite{bornea2024telcorag}. In the following, if not indicated differently, Telco-oRAG uses GPT-3.5 as base model in each of its LLM components and its NN router filters k=5 3GPP series to extract domain-specific content from 3GPP Release 18 documents. The accuracy reported in each experiment is measured as the percentage of correct answers.

In our experiments, we have used four \ac{MCQ} datasets constructed from 3GPP technical documents, which allow us to assess Telco-oRAG across both targeted and broad-spectrum question types:

\begin{itemize}
    \item \textbf{Optimization dataset}: A set of 2000 \acp{MCQ} generated using the methodology proposed in~\cite{maatouk2023teleqna}, based on 3GPP Release 18 documents. 

    \item \textbf{Lexicon dataset}: 500 questions focusing on telecom abbreviations and terminology.
    
    \item \textbf{3GPP Standard dataset}: A curated set of 1840 MCQs extracted from the TeleQnA~\cite{maatouk2023teleqna}, targeting 3GPP standard queries.

    \item \textbf{TeleQnA}:  A dataset including 10000 telecom-related MCQs, based on research manuscripts and documents produced by different standard organization~\cite{maatouk2023teleqna}.
\end{itemize}

\subsection{Hyperparameter Optimization}
\label{sec:retrieval_optimization}

In this section, we present experiments on the optimization dataset and the related tuning of the Telco-oRAG hyperparameters. 
\subsubsection{Chunk Size Optimization}
\label{sec:Chunk Size optimization}

The chunk size defines the size (in tokens) of each segmented text chunk processed during retrieval. For a given context length, smaller chunks increase retrieval granularity, but also increase index size and compute requirements.
Table~\ref{tab:embedding_refinement_comparison} reports a comparison of Telco-oRAG performance across two chunk sizes —- 250 and 500 tokens —- compared along two embedding models: \texttt{text-embed-ada-002} and \texttt{text-embed-3-large}. For each configuration, we show the accuracy achieved with one or two retrieval rounds from the 3GPP database.

As discussed in Section \ref{subsec:RAGRAM}, large chunk size, i.e., 500 tokens or more, is the common option as it reduces the memory requirements of the \ac{RAG} database; however, our results show that this comes at the cost of limited accuracy. Indeed, selecting the chunk size of 250 tokens leads to the best performance across both raw and augmented queries as well as different embedding models, which highlights the importance of a fine retrieval granularity to capture relevant information when dealing with telecom domain queries.

The highest observed accuracy is 79.6\%, achieved using 250-token chunks with the \texttt{text-embed-3-large} model and two rounds of retrieval. Notably, both the token configurations exhibit substantial gains from the designed two rounds retrieval, with improvements up to +2.5\% and +3.4\%, for the 250 chunk size and 500 chunk size, respectively.

\begin{table*}[ht]
\caption{Accuracy achieved through different embedding models for different chunk sizes.}
\label{tab:embedding_refinement_comparison}
\centering
\small
\setlength{\tabcolsep}{8pt}
\renewcommand{\arraystretch}{1.3}
\begin{tabular}{@{}lcc>{\raggedleft\arraybackslash}p{2cm}>{\raggedleft\arraybackslash}p{2.5cm}@{}}
\toprule
\multicolumn{1}{c}{\textbf{Embedding model}} & 
\multicolumn{1}{c}{\textbf{Chunk size}} & 
\multicolumn{1}{c}{\textbf{\# of chunks in the context}} & 
\multicolumn{1}{c}{\textbf{Single round retrieval}}
& \multicolumn{1}{c}{\textbf{Dual rounds retrieval}}\\
\midrule


\addlinespace[2pt]

\rowcolor[gray]{0.96}
\multicolumn{5}{@{}l}{\textbf{\color{black}Medium Chunk Configuration}} \\
Text-embed-ada-002  & 250 & 8 & 77.0\% & 79.5\% (\textbf{+2.5\%})    \\
Text-embed-3-large  & 250 & 8 & 78.4\% & 79.6\% (\textbf{+1.2\%}) \\
\rowcolor[gray]{0.88}
\multicolumn{3}{l}{\textit{Avg. Gain across Embedding Models}} &  
\multicolumn{1}{c}{\phantom{0}\textbf{{+1.4\%}\;}} &  \multicolumn{1}{r}{\textbf{{+0.1\%}\phantom{0000000000}}} \\  

\addlinespace[2pt]

\rowcolor[gray]{0.96}
\multicolumn{5}{@{}l}{\textbf{\color{black}Large Chunk Configuration}} \\
Text-embed-ada-002  & 500 & 4 & 74.0\% & 77.4\% (\textbf{+3.4\%})\\
Text-embed-3-large  & 500 & 4 & 77.4\% & 78.8\% (\textbf{+1.4\%})\\
\rowcolor[gray]{0.88}
\multicolumn{3}{l}{\textit{Avg. Gain across Embedding Models}} & 
\multicolumn{1}{c}{\phantom{0}\textbf{{+3.4\%}}} & \multicolumn{1}{r}{\textbf{{+1.4\%}\phantom{0000000000}}} \\
\bottomrule
\end{tabular}

\vspace{6pt}
\end{table*}

\subsubsection{Context Length Optimization}
\begin{figure}[t]
\centering
\includegraphics[width=0.95\linewidth]{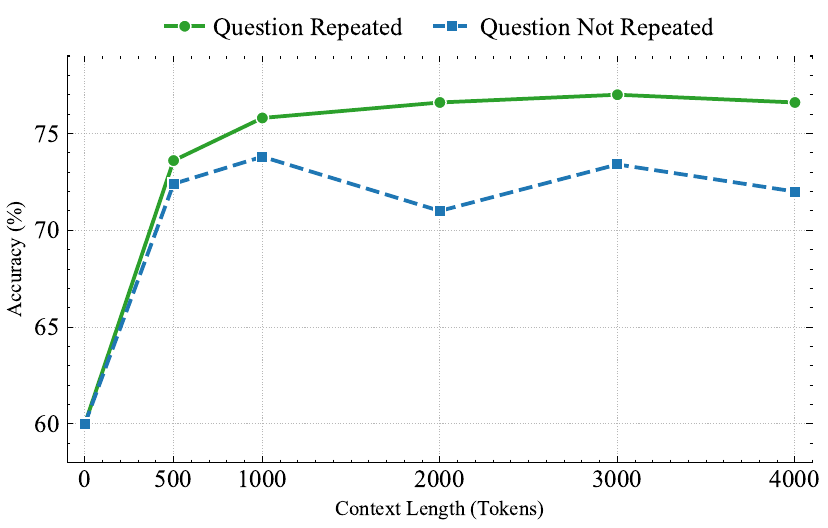}
\caption{RAG accuracy vs. context length.}
\label{fig:increasingcontextlength}
\end{figure}

The context length controls the number of tokens from retrieved documents that are fed into the language model,
influencing both the completeness of the generated answers and the LLM ability to capture long-range dependencies.
In this experiment, we analyze the impact of context length on the answer accuracy by varying the total number of tokens retrieved from the 3GPP corpus. Figure \ref{fig:increasingcontextlength} shows the achieved accuracy as a function of context length, using a fixed chunk size of 250 tokens, and compares two prompt formats: one where the user question appears only once, and another where the question is included both before and after the retrieved context (see Section~\ref{subsec:prompt}).


When the question is presented only once, accuracy declines noticeably beyond 1500 tokens, likely due to attention dilution. In contrast, when the question is repeated before and after the context, accuracy consistently improves as the context length increases up to to 2000 tokens. Beyond this point, adding additional content yields no significant gains.

Based on these findings, we select a context length of 2000 tokens for the remainder of our experiments. Moreover, the results highlight the importance of careful prompt design to maintain LLM performance in long-context settings.


\subsubsection{Embedding Model Selection}
The embedding model converts queries and documents into vector representations. We compare \texttt{text-embedding-ada-002} and \texttt{text-embedding-3-large}, which differ in embedding dimensionality and semantic encoding capacity ~\cite{openai2023embeddings}.
The latter model incorporates Matryoshka Representation Learning~\cite{kusupati2022matryoshka}, which improves performance while compressing vector representations. With a fixed embedding dimension of 1024, \texttt{text-embedding-3-large} consistently outperforms \texttt{text-embedding-ada-002}, achieving an average accuracy improvement of 2.29\% on the optimization dataset. Consequently, we adopt \texttt{text-embedding-3-large} as the default embedding model for Telco-oRAG.

\subsubsection{Indexing Strategy Selection}
We consider three FAISS-based~\cite{johnson2017faiss} similarity search methods: \texttt{IndexFlatIP}, \texttt{IndexFlatL2}, and \texttt{IndexHNSW}. Specifically, \texttt{IndexFlatIP} uses inner product, \texttt{IndexFlatL2} computes exact Euclidean distance, and \texttt{IndexHNSW} provides approximate nearest-neighbor search with sublinear query time.
As mentioned in Section \ref{sec:RAG}, cosine similarity is equivalent to the inner product when using $\ell_2$-normalized embeddings, making both \texttt{IndexFlatIP} and \texttt{IndexFlatL2} theoretically suitable for retrieval. Our empirical results have confirmed this equivalence in practice, with both strategies yielding nearly identical rankings. However, despite only marginal differences in accuracy, IndexFlatIP has outperformed IndexFlatL2 in 80\% of our experiments. We hypothesize that this advantage arises from FAISS optimized handling of inner product computations, which may benefit from lower numerical error accumulation compared to explicit $\ell_2$-distance evaluation. This difference, while subtle, proved consistent across diverse subsets, suggesting that implementation-level details can have a measurable effect even when methods are theoretically equivalent.
By contrast, \texttt{IndexHNSW}, an approximate method optimized for speed, leads to substantial accuracy degradation due to its non-exact nature. Nevertheless, we acknowledge that \texttt{IndexHNSW} may remain attractive in large-scale or latency-sensitive deployments, where its speed offers a practical trade-off between retrieval quality and efficiency.

Accordingly, \texttt{IndexFlatIP} is selected as the default indexing strategy in Telco-oRAG.

\subsubsection{Prompt Engineering}
We conclude this subsection, presenting the improvement in accuracy achieved by the enhanced prompt design, detailed in Section \ref{subsec:prompt}.
By restructuring the queries into a conversational format, we have observed an average accuracy improvement of 4.6\%, as compared to the baseline JSON format used in TeleQnA. 


\subsection{Query Augmentation}






\begin{table}[t]
\centering
\caption{Impact of glossary-enhancement on the LLM capability to answer to lexicon-focused question.}
\begin{tabular}{@{}lcc@{}}
\toprule
\textbf{No Context (LLM Only)} & \textbf{Benchmark RAG} & \textbf{Telco-oRAG} \\ 
\midrule
80.2\% & 84.8\% & \textbf{90.8\%} \\ 
\bottomrule
\end{tabular}
\label{tab:lexicones}
\end{table}

Table~\ref{tab:lexicones} compares the performance achieved on the lexicon dataset by three configurations:
(i) a no-context baseline, where the LLM answers each question without access to external documents, (ii) the ``Benchmark RAG'' without query refinement, and (iii) Telco-oRAG, which includes glossary enhancement. The results show that Telco-oRAG achieves 90.8\% accuracy, improving over the ``Benchmark RAG'' (84.8\%) by +6.0\%, and over the no-context baseline (80.2\%) by +10.6\%. These findings demonstrate that glossary-enhanced queries significantly support the LLM ability to resolve abbreviations and domain-specific terminology.

\begin{figure}[t]
\centering
\includegraphics[width=0.85\linewidth]{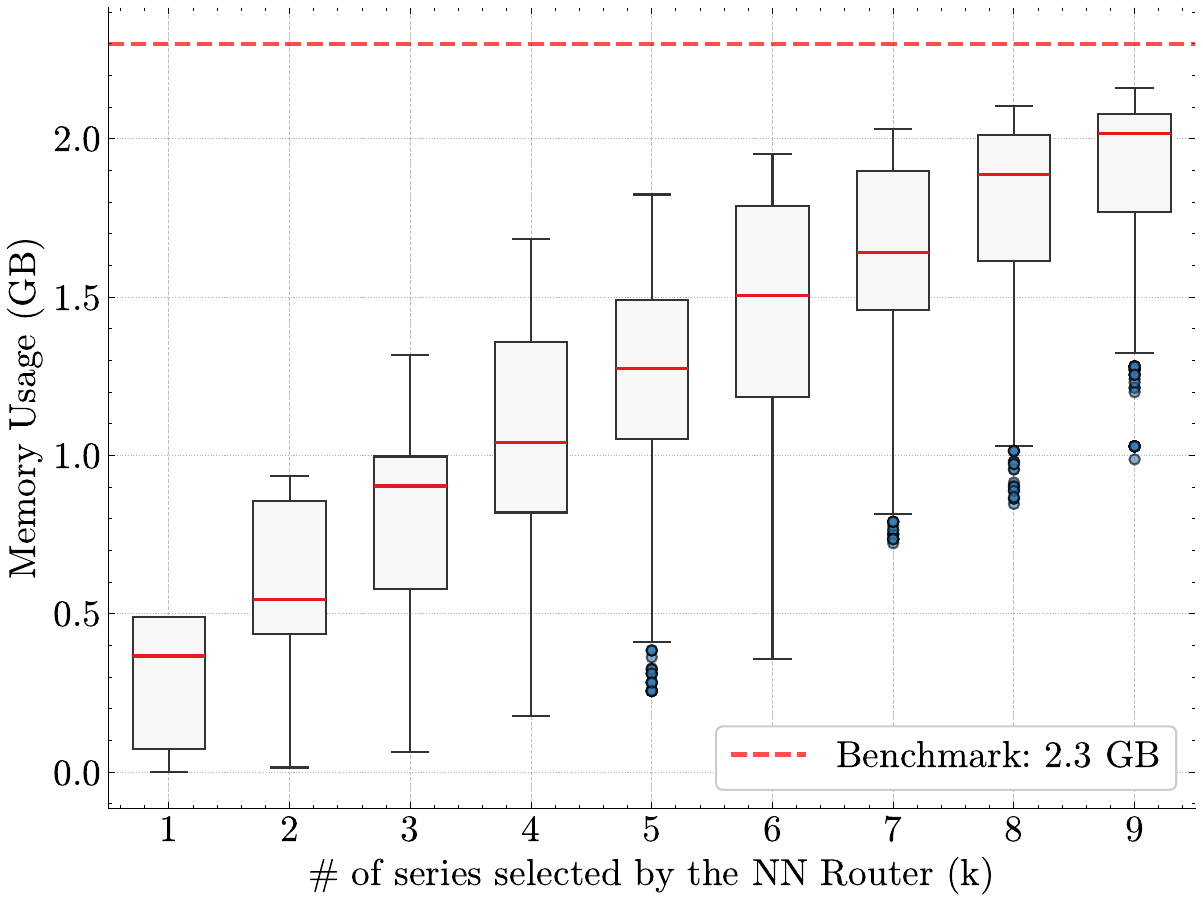}
\caption{Boxplot of RAM usage across 300 sampled queries for Telco-oRAG.}
\label{fig:memory}
\vspace{-1em}
\end{figure}

\subsection{The Memory Footprint of Telco-oRAG}
As shown in Table \ref{tab:embedding_refinement_comparison}, reducing the chunk size improves the model accuracy; however, as discussed in Section~\ref{subsec:RAGRAM}, this also significantly increases the memory footprint of RAG due to the large number of chunks processed per query. 
Figure~\ref{fig:memory} presents a boxplot of the Telco-oRAG RAM usage over 300 sampled queries by varying the number of 3GPP series selected by the NN router. Telco-oRAG leverages the designed NN router capabilities to filter the most relevant sources of information, which leads to a median RAM consumption of 1.25 GB, in contrast to 2.3 GB for a Benchmark RAG, without NN router. This result represents a 45\% reduction in memory usage. 

To further evaluate the NN router retrieval performance, we benchmark its top-$k$ accuracy against two LLM-based baselines (GPT-3.5 and GPT-4o) and a classical \textit{k}-nearest neighbors (\textit{k}-NN) method. Importantly, we also compare three versions of the NN router: the standard one that uses both input streams and with optimized parameters $\alpha$ and  $\beta$. The version without the raw query vector ($\alpha = 0 $) and the one without the alignment-based scoring vector ($\beta = 0 $).
Each model is tasked with predicting the most relevant 3GPP series for a given query, framed as a multi-label classification problem. Top-$k$ accuracy is defined as the percentage of queries for which the ground-truth series appears among the top-$k$ predictions.

\begin{table}[t]
\centering
\caption{Evaluation of the Top-$k$ retrieval accuracy.}
\label{tab:NN_compare}
\small
\renewcommand{\arraystretch}{1.3}

\resizebox{0.95\linewidth}{!}{%
\begin{tabular}{@{}lccccc@{}}
\toprule
\multicolumn{1}{c}{\textbf{Model}} &
\multicolumn{1}{c}{\textbf{Top-1}} &
\multicolumn{1}{c}{\textbf{Top-2}} &
\multicolumn{1}{c}{\textbf{Top-3}} &
\multicolumn{1}{c}{\textbf{Top-5}} &
\multicolumn{1}{c}{\textbf{Top-9}} \\
\midrule

\rowcolor[gray]{0.96}
\multicolumn{6}{@{}l}{\textbf{\color{black}Neural Router}} \\
NN Router & \textbf{51.3\%} & \textbf{71.2\%} & \textbf{80.6\%} & \textbf{88.3\%} & \textbf{97.0\%} \\
NN Router ($\alpha=0$) & 50.3\% & 69.8\% & 78.4\% & 86.3\% & 95.6\% \\
NN Router ($\beta=0$)  & 29.6\% & 50.6\% & 63.1\% & 76.6\% & 95.0\% \\

\addlinespace[2pt]

\rowcolor[gray]{0.96}
\multicolumn{6}{@{}l}{\textbf{\color{black}Baselines}} \\
GPT-3.5 & 19.9\% & 27.6\% & 36.6\% & 50.3\% & 78.2\% \\
GPT-4o  & 30.4\% & 56.2\% & 70.8\% & 85.6\% & 89.7\% \\
\textit{k}-NN & 15.3\% & 24.3\% & 29.8\% & 42.3\% & 57.0\% \\
\bottomrule
\end{tabular}%
} 
\vspace{-0.5em}
\end{table}

\begin{figure}[t]
\centering
\includegraphics[width=1\linewidth]{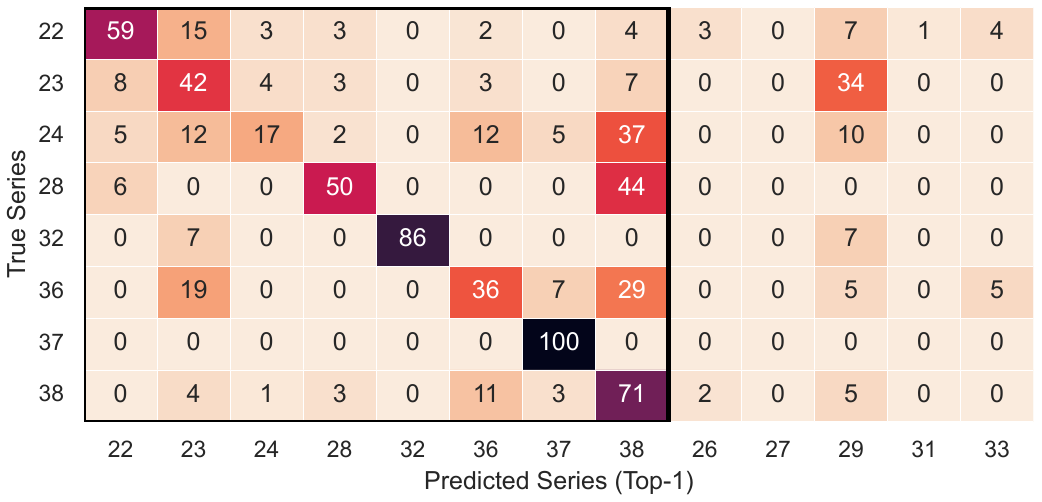}
\vspace{-1em}
\caption{3GPP series confusion matrix (row-normalized \%).}
\label{fig:memory_a}
\vspace{1em}
\end{figure}

As we observe in Table~\ref{tab:NN_compare}, the standard NN router achieves a top-3 accuracy of 80.6\%, outperforming GPT-4o (70.8\%) and GPT-3.5 (36.6\%) by 9.8\% and 44.0\%, respectively. However, when ablating one of the two input streams, performance drops substantially. In particular, the loss of the alignment signal ($\beta = 0 $) leads to the largest degradation, highlighting the importance of coarse alignment between the query and the 3GPP series summaries.

\begin{figure}[t]
\centering
\includegraphics[width=0.8\linewidth]{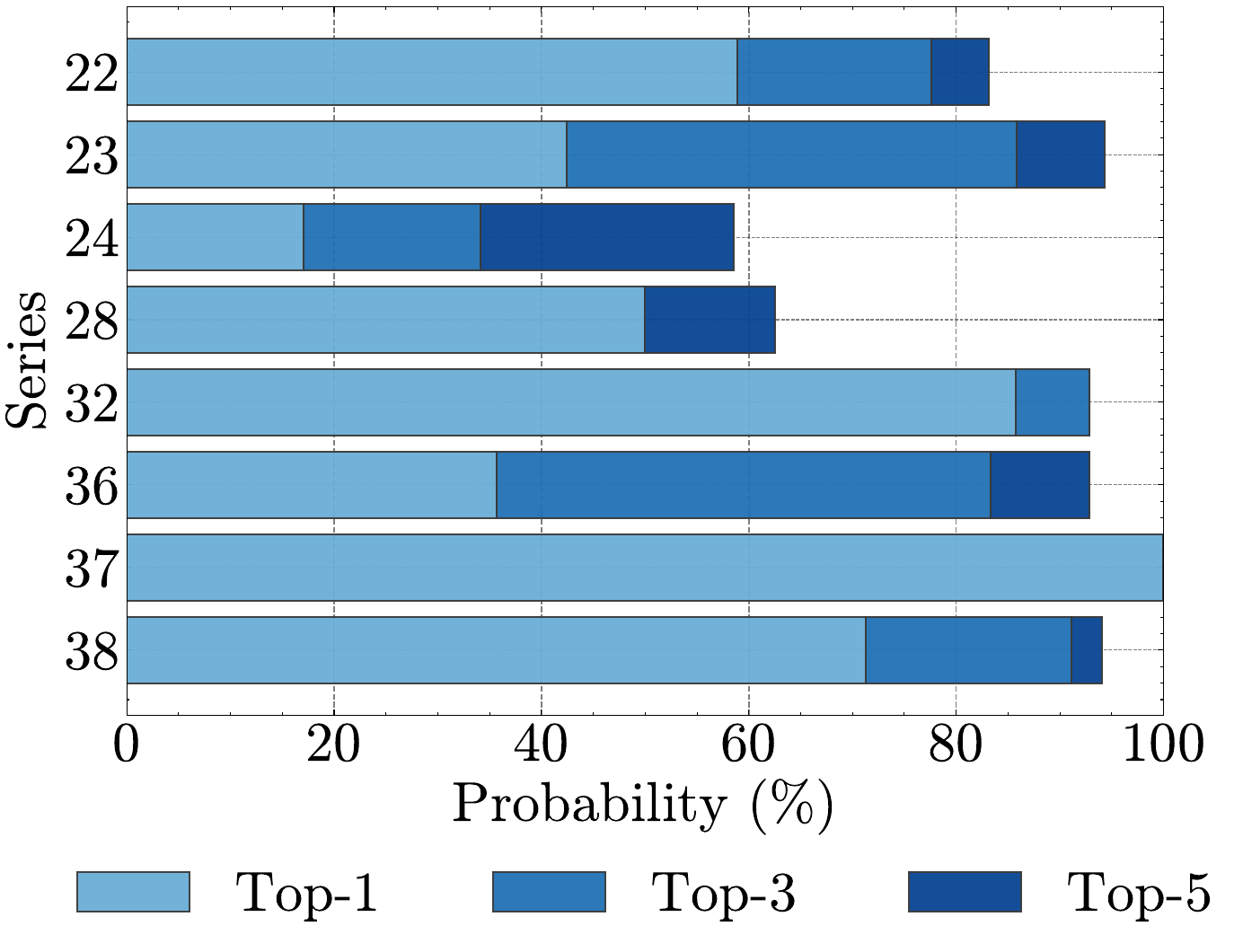}
\vspace{-1em}
\caption{NN router prediction quality per 3GPP series.}
\label{fig:memory_b}
\vspace{-1em}
\end{figure}

Figure~\ref{fig:memory_a} the distribution of the misclassified questions across 3GPP series. These results show that the errors exhibit some patterns such as 24 series and 28 series being frequently confused with 38 series or 23 series with 29 series. These series operate in adjacent or semantically related domains within 3GPP, often sharing overlapping technical terminology and collaborative work items.

In addition, Figure 4 reports the per–working-group prediction accuracy at Top-1, Top-3, and Top-5 levels. The results demonstrate that while several working groups achieve high reliability at Top-1, others exhibit substantially lower performance. Notably, a 50\% accuracy threshold across working groups is only attained at the Top-5 level, thereby motivating our choice to propagate a Top-5 shortlist downstream in the pipeline. Finally, despite these misclassifications, our experiments (using GPT-5-mini) show that Telco-oRAG performance does not drop with respect a baseline model, even when all the 5 retrieved documents are wrong, highlight that our pipeline remain robust, thanks to the web retrieval stage.

\subsection{Is Online All You Need?}
\label{subsec:onlineonly}

\begin{figure}[ht]
\centering
\includegraphics[width=0.95\linewidth]{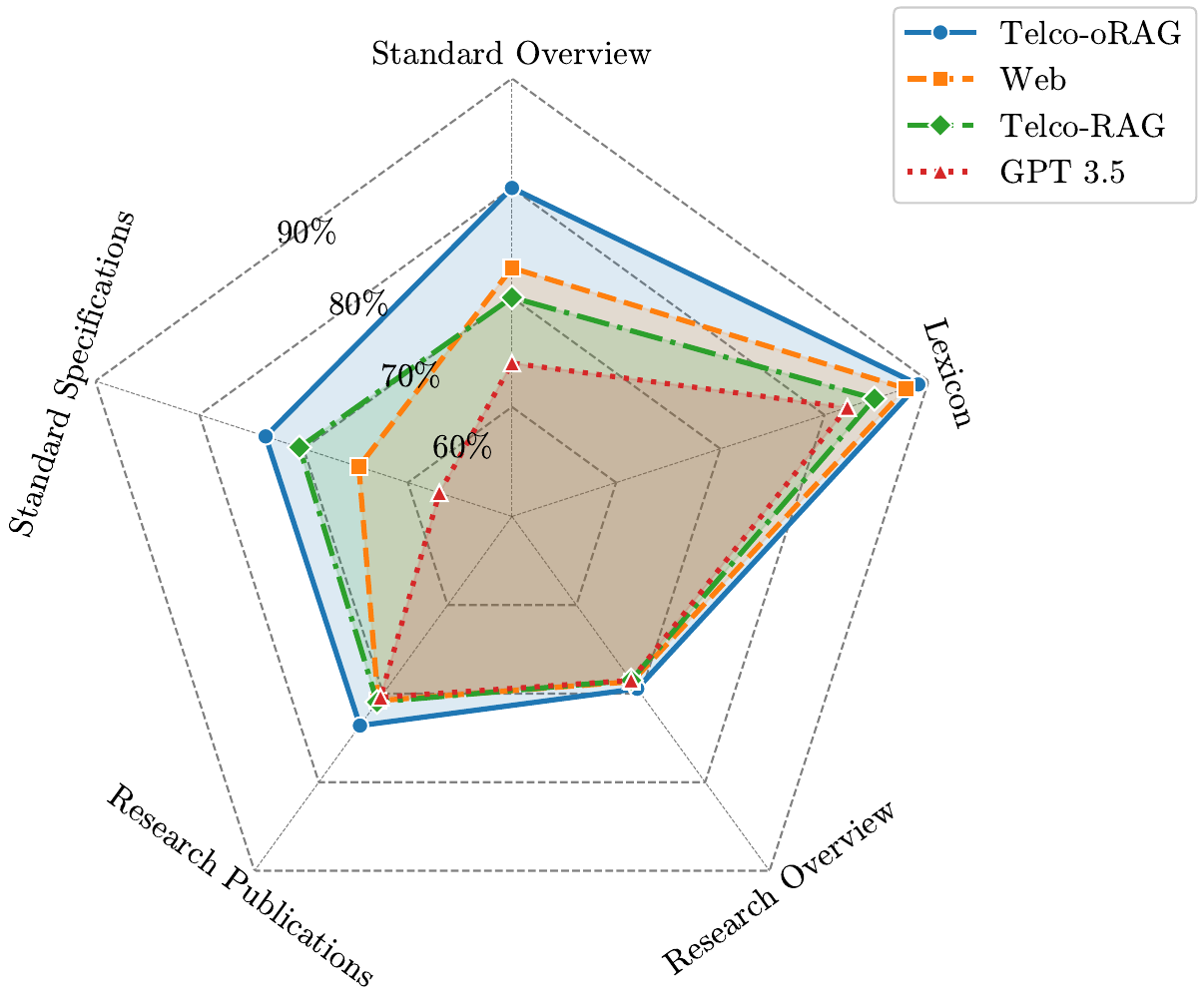}
\vspace{-1em}
\caption{Accuracy of different models on TeleQnA.}
\label{fig:Telco-Online}
\end{figure}

In this section, we compare web search and classic \ac{RAG} with respect to different types of telecom questions and analyze the gain brought by combining them together, i.e., in Telco-oRAG.
In Figure~\ref{fig:Telco-Online}, using TeleQnA, we compare four LLM configurations based on GPT-3.5:
\begin{enumerate}
    \item \textbf{GPT-3.5};
    \item \textbf{Web}: GPT-3.5 with web search;
    \item \textbf{Telco-RAG} \cite{bornea2024telcorag};
    \item \textbf{Telco-oRAG}.
\end{enumerate}

These results highlight that the baseline GPT-3.5 model achieves limited accuracy, in particular for standard overview and standard specification questions. However, both the web retrieval and Telco-RAG significantly enhance the baseline model performance in these domains as well as for the lexicon MCQs. Notably, the online retrieval outperforms Telco-RAG on standard overview while Telco-RAG achieves better accuracy than the online retrieval on standard specifications. These results are due to the limited capability of Telco-RAG to answer to questions related to non-3GPP standards, as its embedding database is based on 3GPP Rel.18 documents. Finally, Figure~\ref{fig:Telco-Online} shows that Telco-oRAG---combining domain-specific and online retrieval---achieves the highest accuracy across all TeleQnA categories, most notably on "Standards Overview" where it brings $\approx$10 pp in comparison to Telco-RAG.



\begin{figure}[t]
\centering
\includegraphics[width=1\linewidth]{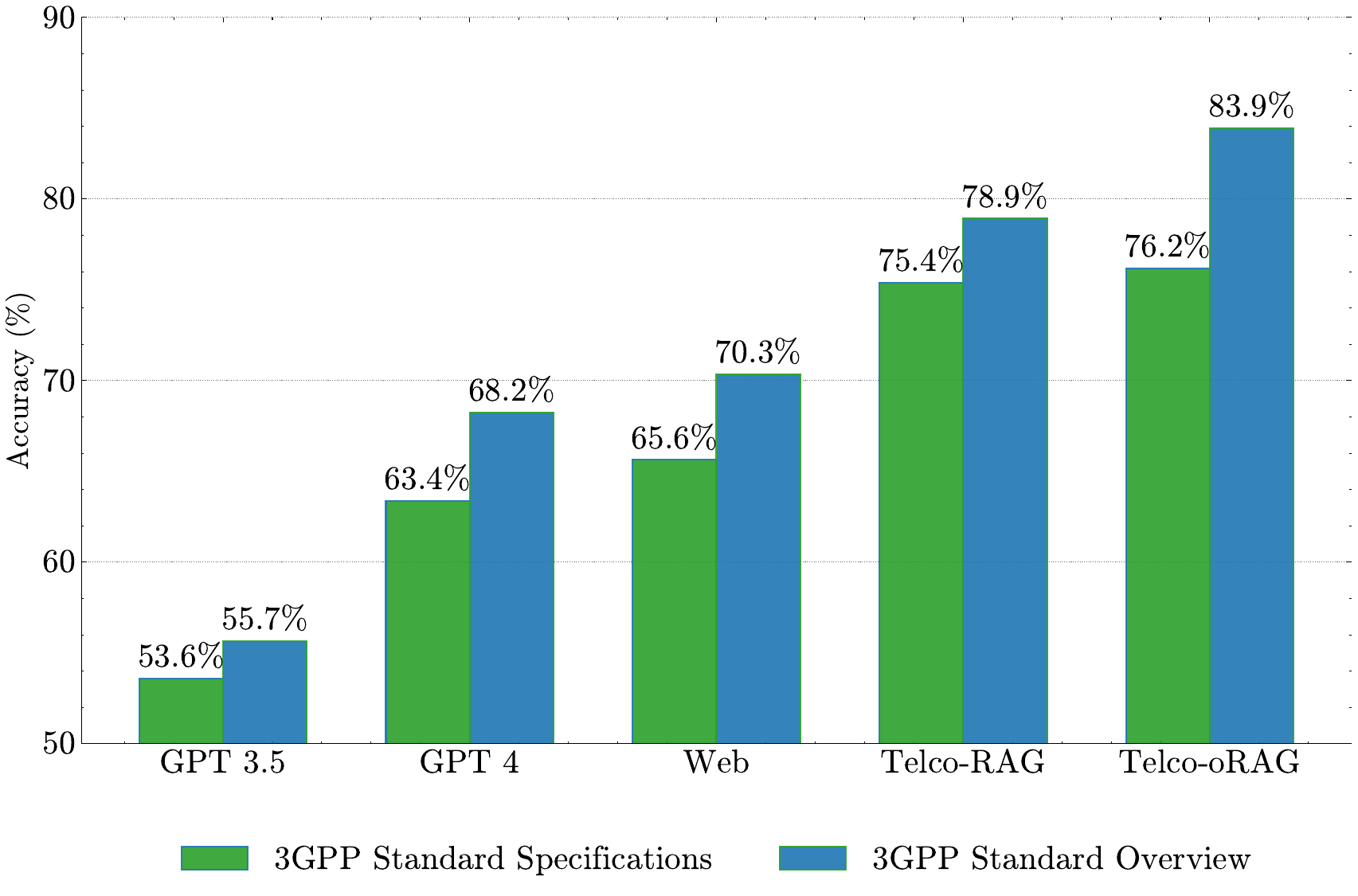}
\vspace{-1em}
\caption{Accuracy of different models on 3GPP Standard dataset.}
\label{fig:OnlineRetrieval}
\vspace{-1em}
\end{figure}

Figure~\ref{fig:OnlineRetrieval} provides detailed results of different models focusing specifically on the 3GPP Standard datasets, which includes only 3GPP Standard Specifications and 3GPP Standard Overview \acp{MCQ}. 
In addition to the models reported in Figure~\ref{fig:Telco-Online}, Figure~\ref{fig:OnlineRetrieval} also includes the performance achieved by GPT-4 without retrieval.\footnote{We did not include GPT-4 in Figure~\ref{fig:Telco-Online} for readability purposes.}


On {3GPP Standard Specifications} \acp{MCQ}, GPT-3.5 and GPT-4 achieve 53.6\% and 63.4\% of accuracy, respectively. Integrating web search in GPT-3.5 raises its performance to 65.6\%. Telco-RAG significantly outperforms these models with 75.4\% accuracy, underscoring the benefit of structured and domain-specific retrieval for answering to questions related to highly technical documents. However, Telco-oRAG  achieves the best performance (76.2\%), leading to more than 20\% and 10\% of improvement with respect to GPT-3.5 and GPT-4.

On {3GPP Standard Overview} \acp{MCQ}, GPT-3.5 and GPT-4 achieve 55.7\% and 68.2\% of accuracy, respectively. Augmenting GPT-3.5 with web search increases accuracy its 70.3\%. Telco-RAG again outperforms the other baseline model, achieving 78.9\% of accuracy. Finally, Telco-oRAG provides the best performance with 83.9\% of accuracy, leading to more than 25\% and 15\% improvement over GPT-3.5 and GPT-4. 
z
It is important to note that the performances in Figure~\ref{fig:OnlineRetrieval} are not the same as the one for the standard specifications and standard overview in Figure~\ref{fig:OnlineRetrieval}, as 1) TeleQnA includes \acp{MCQ} from different standards while the 3GPP evaluation dataset has questions only related to 3GPP documents, 2) in our experiments, Telco-RAG and Telco-oRAG leverage a database composed exclusively by 3GPP Rel. 18 documents, which makes them particularly effective on the 3GPP evaluation dataset.

\subsection{Is Telco-oRAG Architecture LLM-dependent?}

To assess how Telco-oRAG architecture generalizes across different LLMs, we evaluated its performance using various models, including GPT-3.5, LLAMA-3-8B, LLAMA-3-70B, Mistral-7B, and Qwen-72B. The motivation behind this analysis is to determine how well Telco-oRAG enhances the accuracy of different LLMs, particularly in answering MCQs related to 3GPP standard documents.

\begin{figure}[ht]
\centering
\includegraphics[width=1\linewidth]{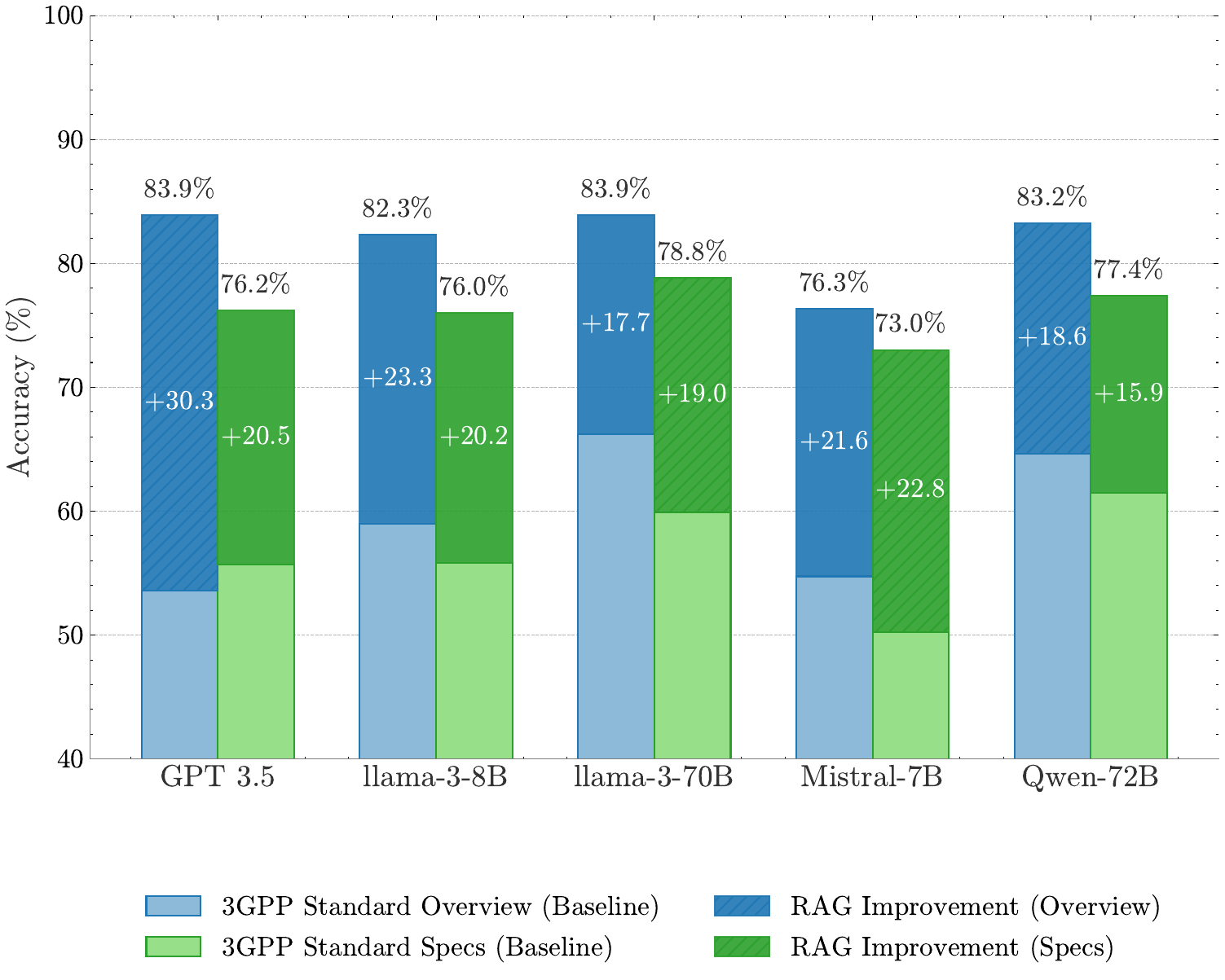}
\vspace{-1em}
\caption{Comparative accuracy of Telco-oRAG combined with various LLMs on 3GPP Standard dataset. }
\label{fig:modelsgain}
\end{figure}

Figure~\ref{fig:modelsgain} presents the comparative accuracy of Telco-oRAG with respect to the baseline performance of each vanilla model, when used to answer to \acp{MCQ} from the 3GPP Standard Overview and 3GPP Standard Specifications datasets. 
GPT-3.5 exhibits the highest relative improvement: a gain of +30.3\% in 3GPP Standard Overview (from 53.6\% to 83.9\%) and +20.5\% in the Standard Specification (from 55.7\% to 76.2\%).

Importantly, the results presented in Figure~\ref{fig:modelsgain} indicate substantial performance improvements in all models when utilizing Telco-oRAG, with an average gain of 19.7\% and 22.3\% in answering questions about 3GPP Standard Specifications 3GPP Standard Overview, respectively. These findings suggest that Telco-oRAG can be effectively adapted to work with various LLMs to answer questions related to telecom standards. 



These findings suggest that while the magnitude of improvement varies across different LLMs, the consistent accuracy gains observed confirm that Telco-oRAG can be effectively adapted to work with various LLMs, enhancing their ability to handle telecom-specific questions.

\begin{figure*}[ht!]
\centering
\includegraphics[width=1\linewidth]{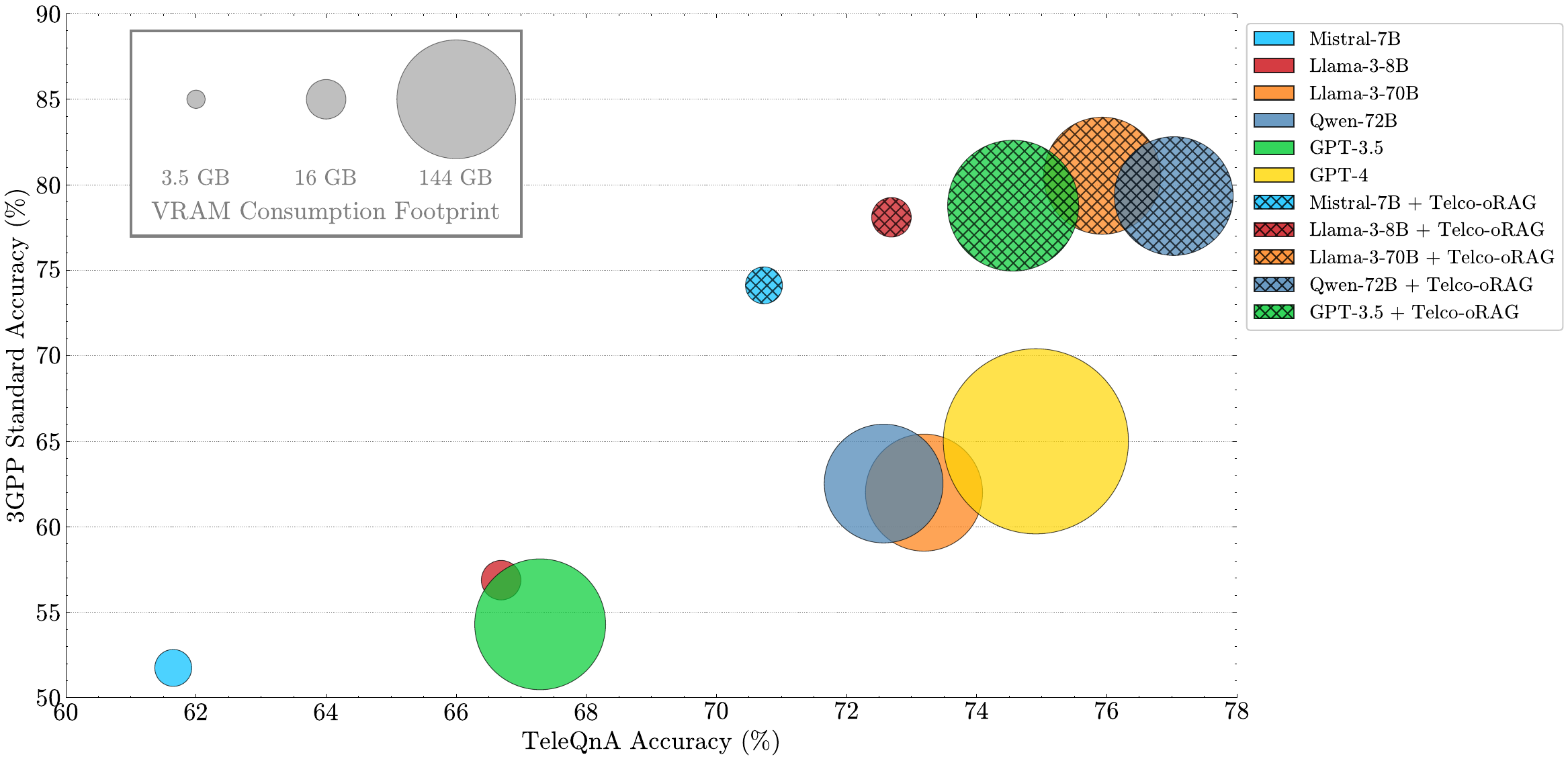}
\vspace{-1em}
\caption{Plotted performance of language models with and without Telco-oRAG, where the size of the circle signifies memory footprint.}
\label{fig:vram}
\vspace{-1em}
\end{figure*}

\subsection{LLMs for Memory-constrained Devices}
\label{subsec:SLMvsLLM}

In some use cases, the choice of \ac{LLM} depends not only on the performance but also on the memory requirements; this can be the case where \acp{LLM} are deployed on mobile devices or edge cloud servers.
In these use cases \acp{LLM} such as LLAMA-3-8B and Mistral-7B, or quantized models, offer a lightweight alternative to resource-hungry \ac{LLM} such as GPT-4; however, these resource-efficient solutions
typically underperform in complex, domain-specific tasks. In the following experiments, we assess the ability of Telco-oRAG to mitigate this limitation and help resource-efficient models to achieve performance levels previously reserved for resource-hungry models.

In Figure~\ref{fig:vram}, we compare the performance of vanilla, quantized, and Telco-oRAG models in terms of accuracy on the {3GPP Standard} and TeleQnA datasets. For each model represented in Figure~\ref{fig:vram}, its circle size indicates its memory footprint. Telco-oRAG enables Mistral-7B (14~GB VRAM) to achieve 74.12\% on 3GPP Standard and 70.73\% on the TeleQnA, substantially outperforming GPT-3.5 (54.3\% and 67.29\%) and providing better performance than GPT-4 (65.0\% and 74.91\%) on 3GPP Standard at a fraction of its resource cost. Similarly, Llama-3-8B combined with Telco-oRAG achieves 78.10\% on 3GPP Standard, further highlighting the effectiveness of Telco-oRAG in closing the performance gap between mid-sized \acp{LLM} and proprietary LLMs on domain-specific knowledge.

\subsection{Benchmarking on 3GPP Releases~18 questions}

To conclude our analysis we compare in Table~\ref{tab:accuracy_comparison} the performance of Telco-oRAG with other telecom-specialized models on 3GPP Releases~18 questions from the 3GPP Standard dataset. We use the following models as benchmark:
\begin{enumerate}
    \item \textbf{LLama-3-8B-Tele-it}: A model based on LLama-3-8B and specialized in telecommunications using using \ac{SFT} \cite{maatouk2024telellmsseriesspecializedlarge};
    \item \textbf{Chat3GPP}: A model based on LLama3-8B-Instruct that integrates an open \ac{RAG} framework \cite{huang2025chat3gpp};
    \item \textbf{Telco-RAG}: the RAG model presented in \cite{bornea2024telcorag}, without web search.
\end{enumerate}

Note that Chat3GPP, Telco-RAG, and Telco-oRAG include in their database 3GPP Releases~18 documents. In contrast, LLama-3-8B-Tele-it has been fine-tuned with 2.8k 3GPP documents from different releases.
Our results confirm that Telco-oRAG achieves the highest accuracy (80.9\%), outperforming both other \ac{RAG} models, Telco-RAG (78.4\%) and Chat3GPP (79.1\%), as well as a fine-tuned model, LLama-3-8B-Tele-it (57.1\%). 



\begin{table}[htbp]
\centering
\caption{Accuracy Comparison of telecom-specialized models on 3GPP Release~18 questions.}
\label{tab:accuracy_comparison}
\begin{tabular}{lcc}
\toprule
\textbf{Model} & \textbf{Rel.18} \\
\midrule
LLama3-8B-Tele-it & 0.571\\
Chat3GPP & 0.791 \\
Telco-RAG & 0.784 \\
\rowcolor{gray!20}
\textbf{Telco-oRAG}  & 0.809  
\\
\bottomrule
\end{tabular}

\vspace{11pt}
\end{table}

\subsection{Evaluating Telco-oRAG on open-ended questions}

We have evaluated how Telco-oRAG performs on 250 open-ended questions from TeleQnA using an LLM-as-Judge approach and we have compared its performance against a baseline solution without RAG. 
As a reference, we have also provided the accuracy computed on the MCQ versions of the same questions. Judges make each evaluation based on the question, its ground truth, and the provided answer.

Across all the experiments presented in Table~\ref{tab:accuracy_comparison_OE} Telco-oRAG consistently outperforms the baseline on open-ended evaluation: the gains range from +27.3 to +42.8 percentage points. In the MCQ scoring, GPT-5-mini and GPT-OSS-120B achieve close performance. In contrast, we observe that the judges tend to provide higher scores to answers coming from the same model (or from the same family of models). This means that scoring open questions through LLM-as-a-judge does not provide consistent and unbiased results, which makes MCQ evaluation preferable. 

    \begin{table*}[htbp]
    \centering
     \caption{Accuracy comparison between Telco-oRAG and a baseline system (vanilla model without RAG). 
The evaluation is conducted on 250 questions.
}
    \label{tab:accuracy_comparison_OE}
    \begin{tabular}{llcccc}
    \toprule\textbf{}
    \textbf{Generating Model} & \textbf{System} & \multicolumn{3}{c}{\textbf{LLM-as-a-Judge}} & \textbf{MCQ} \\
    \cmidrule(lr){3-5} \cmidrule(l){6-6}
     & & \textbf{Qwen3-32B} & \textbf{GPT-5-mini} & \textbf{GPT-OSS-120B} & \textbf{Accuracy} \\
    \midrule
    \multirow{2}{*}{Qwen3-32B} & Telco-oRAG & 78.2\% & 74.9\% & 58.8\% & 82.4\% \\
                                         & Baseline   & 37.4\%     & 32.1\%     & 17.2\%    & 56.0\% \\
    \cmidrule(lr){1-6}
    \multirow{2}{*}{GPT-5-mini} & Telco-oRAG & 74.9\% & 79.1\% & 67.4\% & 84.3\% \\
                                  & Baseline   & 46.6\%     & 50.9\% & 40.1\%     & 73.0\% \\
    \cmidrule(lr){1-6}
    \multirow{2}{*}{GPT-OSS-120B} & Telco-oRAG & 74.4\% & 76.7\% & 63.5\% & 84.7\% \\
                                            & Baseline   & 42.6\%    & 40.4\%     & 32.0\%     & 68.4\% \\
    \bottomrule
    \end{tabular}
    \end{table*}

Table~\ref{tab:qualitative} shows a side-by-side evaluation of
answers produced by Telco-oRAG (using GPT-5-mini), Telco-RAG (using GPT-5-mini), and
TelecomGPT [1] on an open-ended question derived from TeleQnA. 
The evaluation of the
response accuracy is realized using human evaluation. 
In this example,the context of the query lies outside the embedding databases available for Telco-RAG and Telco-oRAG. TelecomGPT does not use a RAG mechanism to provide specific context to the LLM generating the answer to the user's query. Due to these facts, both TelecomGPT and Telco-RAG produce responses that seem related to the query but are incorrect.
In contrast, Telco-oRAG provides the correct answer by leveraging the web retrieval output.
   
 \begin{table*}[h]
    \centering
    \caption{Qualitative Comparison of RAG Systems.}
    \label{tab:qualitative}
    \begin{tabular}{p{2cm} p{11cm} p{1.5cm}}
    \toprule
    
    \textbf{Question} & \textbf{Which protocol is associated with port number 2222? [TCP/IP]} & \\
    
    \midrule
    
    \textbf{System} & \textbf{Answer Provided} & \textbf{Evaluation} \\
    
    \midrule
    Telco-RAG & ... Generally, port 2222 is commonly used by some Secure Shell (SSH) servers, particularly as an alternative to the standard SSH port 22. However, its usage can vary widely based on specific server configurations and application needs. Without specific information from the retrievals linking port 2222 to a distinct protocol within the context of the referenced documents... & Wrong \\
    \addlinespace

    TelecomGPT & The port number 2222 is not officially assigned to any specific protocol by the Internet Assigned Numbers Authority (IANA). However, it is often used for various purposes in different applications and services. For example, it can be used for:

    IRCd (Internet Relay Chat Daemon): Some IRC servers use port 2222.
    
    Custom Applications: Many custom or proprietary applications use this port for their specific communication needs. & Wrong \\
    
    \addlinespace
    
    Telco-oRAG &... Based on the IANA Service Name and Port Number Registry 2222 is officially registered for\textbf{ EtherNet/IP-1}. In operational practice it is often used as an alternate SSH port or by vendor-specific application... & Correct \\

    \addlinespace
    \multicolumn{3}{c}{} \\ 
    \midrule
    \textbf{Ground Truth} & \textbf{The port number 2222 is associated with EtherNet/IP protocol.} & \\
    \bottomrule
    \end{tabular}
    \end{table*}

    \subsection{Latency-performance tradeoff in Telco-oRAG}
\label{subsec:latency_analysis}

An important concern regarding multi-stage retrieval–generation pipelines is the potential overhead in inference latency.  To address this concern, we conducted a detailed latency analysis of Telco-oRAG, decomposing end-to-end inference time into its constituent blocks:  
(i) LLM rephrasing, 
(ii) glossary enhancement, 
(iii) web retrieval, 
(iv) 3GPP retrieval, and 
(v) prompt engineering, and (vi) final LLM generation. 

\begin{figure}[H]
      \centering
    \includegraphics[width=0.9\linewidth]{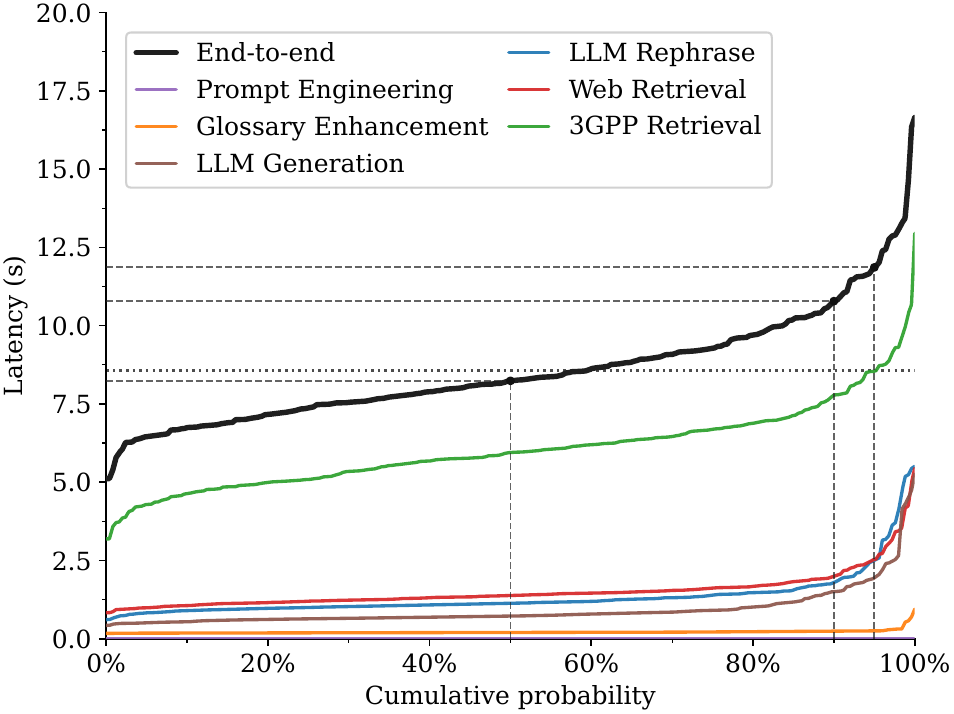}
      \caption{Empirical cumulative distribution function of Telco-oRAG latency across different constituent modules. The dashed lines represent the probability quantiles at the $50\%$, $90\%$, and $95\%$ levels, while the dotted line denotes the mean end-to-end latency.}
        \label{fig:latency1}
    \end{figure}

Figure~\ref{fig:latency1} presents the empirical cumulative distribution functions (ECDFs) for each stage across 500 TeleQnA queries, alongside the total pipeline latency. Despite the multi-stage retrieval design of Telco-oRAG, most of its components have a limited contribution to the overall latency.
In the following, we provide an analysis of the performance-accuracy tradeoff, for each component of Telco-oRAG, to justify its necessity in the overall pipeline: 
\begin{itemize}
\item The Query Refinement block is composed of two components: LLM Rephrasing and Glossary Enhancement.
\begin{itemize}
\item The LLM Rephrasing corrects typos and grammar mistakes in the user question; prior work shows that correcting these errors substantially improves overall RAG performance \cite{zhang2025qe-rag}. LLM Rephrasing accounts for 15.7\% of the end-to-end latency.
\item The Glossary enhancement helps the pipeline to process telecom acronyms and synonyms. As shown in Table II, on the Lexicon
subset of TeleQnA, Telco-oRAG, thanks to this feature, achieves 90.8\% of accuracy, improving over Benchmark
RAG by +6.0 pp and over no-context LLM by +10.6 pp. The glossary enhancement accounts for only 2.3\% of the total latency.
\end{itemize}
\item The web retrieval supplies the LLM supplementary context not available in the embedding (as explained in Section
IV-D). We observe that Telco-oRAG improves the accuracy of Telco-RAG on Standards Overview question from TeleQnA by 10 percentage points (see Fig. 10). Moreover, Table VI highlights that the web retrieval makes Telco-oRAG capable of answering questions related to different standards, whose documents may not be available in the embedding database.  Finally, the web retrieval is done in parallel with the 3GPP Retrieval, therefore it does not increase the end-to-end Telco-oRAG latency.
\item The NN router improves the LLM generation efficiency (as discussed in Section IV-C). Retrieving 
only the top-k 3GPP series reduces the median RAM consumption from 2.3 GB to 1.25 GB (approx. 45\%) with respect to a baseline RAG, without affecting the pipeline accuracy
relevance. The NN Router is a component of 3GPP Retrieval and it accounts for 8.5\% of the overall latency.
\item {LLM Generation} produces the final synthesized answer from the refined query and retrieved context. Using the OpenAI inference, this stage contributes 10.8\% of the total end-to-end latency.
\end{itemize}
To conclude, the dominant cost is the 3GPP retrieval step which accounts for 71\%. Nevertheless, 75–80\% of the perceived latency in the 3GPP retrieval component stems from LLM inference used to generate candidate answers between the two retrieval rounds.
To match the performance of the released pipeline, we evaluated Telco-oRAG latency using LLM inference from OpenAI API. Accordingly, the user experienced latency during LLM rephrasing, 3GPP retrieval, and final LLM generation can be optimised using an open-source LLM deployed at the user compute infrastructure.

\section{Conclusions}\label{s:conclusion}
This paper introduced Telco-oRAG, a modular RAG framework tailored to address the specific challenges of telecom-standard question answering. Going beyond previous work, \mbox{Telco-oRAG} combines query refinement stages, web search, and retrieval from 3GPP specifications orchestrated through a neural router that significantly reduce resource requirements.

We provided a systematic analysis of critical RAG hyperparameters, such as chunk size, context length, embedding model, and prompt formatting, demonstrating their impact on accuracy. The proposed framework proved particularly effective, improving response accuracy on lexicon queries by 10.6\% and overall MCQ accuracy by up to 17.6\% compared to baselines. Additionally, Telco-oRAG achieves a 45\% reduction in VRAM consumption via selective loading of domain-relevant content, enabling deployment on resource-constrained devices.

Our experiments show that Telco-oRAG generalizes across multiple LLMs, including 
open-source models, narrowing the performance gap with proprietary LLMs like GPT-4 while requiring an order of magnitude less memory. Moreover, by integrating real-time web retrieval, Telco-oRAG adapts to evolving standards, surpassing both static RAG pipelines and fine-tuned domain-specific models on current 3GPP content.

By making Telco-oRAG publicly available, we aim to provide a practical foundation for the integration of LLMs into real-world telecom applications. Future works will focus on integrating multimodal capabilities in Telco-oRAG, which will allow processing tables and figures to further enhance its accuracy as well its impact in novel use cases.

\begin{acronym}[AAAAAAAAA]
 \acro{3GPP}{third generation partnership project}
 \acro{AI}{artificial intelligence}
 \acro{API}{application programming interface}
 \acro{BS}{base station}
 \acro{BERT}{bidirectional encoder representations from transformers}
 \acro{DPO}{direct Preference Optimization}
 \acro{FPGA}{field-programmable gate array}
 \acro{GPT}{generative pre-trained transformer}
 \acro{IEEE}{institute of electrical and electronics engineers}
 \acro{ITU}{international Telecommunication Union}
 \acro{GPU}{graphical processing unit}
 \acro{KG}{Knowledge Graph}
 \acro{LLM}{large language model}
 \acro{MAPE}{mean absolute percentage error}
 \acro{MCQ}{multiple-choice question}
 \acro{NN}{neural Network}
 \acro{RAM}{random-access memory}
 \acro{RAG}{retrieval-augmented generation}
 \acro{RAN}{radio access network}
 \acro{RLHF}{reinforcement learning with human feedback}
 \acro{SFT}{supervised fine-tuning}
 \acro{SLM}{small language model}
 \acro{ML}{Machine Learning}
 \end{acronym}

\bibliographystyle{IEEEtran}
\bibliography{reference.bib}

\end{document}